\documentclass{pinchcr}   % Comment the above and uncomment this
\usepackage{graphicx}        % standard LaTeX graphics tool
\usepackage{amsmath,amsfonts,amssymb,latexsym}

\def\dd{{\rm d}}
\def\s{\sigma}
\def\us{\underline{\sigma}}
\def\l{\ell}
\def\lc{\ell_{\rm c}}
\def\t{\tau}
\def\tc{\tau_{\rm c}}

\def\da{{\partial a}}
\def\di{{\partial i}}
\def\dami{{\partial a \setminus i}}
\def\dima{{\partial i \setminus a}}
\def\ve{\varepsilon}
\def\la{\langle}
\def\ra{\rangle}
\def\Bbar{{\bar{B}}}

\def\P{\mathbb{P}}

\title{Analytical approaches to time and length scales in models of glasses}

\author{Silvio Franz}
\affiliation{Universit\'e  Paris Sud, CNRS, LPTMS, UMR8626, Orsay
F-91405, France}

\author{Guilhem Semerjian}
\affiliation{LPTENS, Unit\'e Mixte de Recherche (UMR 8549) du CNRS
et de l'ENS associ\'ee \`a l'universit\'e Pierre et Marie Curie, 24
Rue Lhomond, 75231 Paris Cedex 05, France.}

\begin{document}

\maketitle

\preface
  The goal of this chapter is to review recent analytical results
  about the growth of a (static) correlation length in glassy systems,
  and the connection that can be made between this length scale and
  the equilibrium correlation time of its dynamics. The definition of
  such a length scale is first given in a generic setting, including
  finite-dimensional models, along with rigorous bounds linking it to
  the correlation time. We then present some particular cases (finite
  connectivity mean-field models, and Kac limit of finite dimensional
  systems) where this length can be actually computed.

\section{Introduction}

The ``Random first order theory'' (RFOT) is one of the most widely
discussed theory of glass formation and glassy phenomena in fragile
systems. Its origin is based upon the observation by Kirkpatrick,
Wolynes and Thirumalai~\shortcite{KTW} that a family of abstract
long range spin glass models with ``one step replica symmetry
breaking'' (1RSB)~\shortcite{MPV} presents freezing phenomena that
in several aspects resemble the observed phenomenology in freezing
of fragile liquids and other glassy systems. In this perspective
long range spin glasses provide a unifying description of glassy
phenomena, including dynamical aspects (MCT dynamical singularity
and aging phenomena) and thermodynamical ones (metastability and
Kauzmann like entropy crisis), which have given rise to predictions
verified in simulations and experiments~\shortcite{simexp,Bscience}.
Despite the appeal of the resulting picture, it immediately became
clear that two problems had to be overcome to be able to apply
convincingly the theory to supercooled liquids: 1) The disordered
interactions of spin glasses models is not a realistic microscopic
description of liquid systems 2) The 1RSB picture is strongly
dependent on the long range character of the interactions and
presents some typical pathologies of mean-field theories. Both
problems have attracted a lot of attention.

The first problem has been basically circumvented by the fact that
the most sophisticated Mean-Field theories based on realistic liquid
models of particles in interaction give back the 1RSB scenario
\shortcite{MPgl}. This suggest a high level of universality of
glassy phenomena that goes from spin models with random interactions
to supercooled liquids, one can thus expect to understand general
properties of the latter ones from the study of the former ones,
which are simpler microscopically.

The second problem, that is how to effectively take into account the
finite range character of the interactions, remains in our opinion
the main obstacle against an accomplished theory of the glass
transition, despite important contributions. The RFOT, based on
scaling and phenomenological arguments, proposes an intriguing
scenario about which features of Mean-Field Theory survive in short
range systems and what of the picture should be modified. However, a
theory based on a first principle analysis of microscopic models is
unfortunately lacking.  Understanding the role of finite-range
interactions includes both important questions of principle, but not
directly related to physical observables such as the possibility of
an ideal glass transition in systems with short range interactions,
and questions of direct practical interest as finding a theory for
activated dynamics and the cross-over between mode coupling behavior
and activation.

This chapter reviews some of the theoretical efforts to understand
the relation between the physics of finite dimensional glassy
systems and their mean field description and to include finite range
interactions in microscopic models of glassy phenomena.  A crucial
point that an accomplished theory of the glass transition should
address is the growth of correlations that accompanies the increase
of relaxation time as temperature is decreased.  The relaxation time
of a supercooled liquid increases dramatically upon lowering its
temperature, until reaching experimentally accessible timescales at
the laboratory glass transition temperature $T_{\rm g}$. In this
range of temperature the usual static spatial correlations (for
instance the structure factor measured in scattering experiments)
remain essentially the same as the ones of a high temperature
liquid. These two facts seem contradictory: the physical intuition
relates a large correlation time to cooperative relaxation
mechanisms involving a large volume of the sample (which
incidentally is an argument in favour of the universality of glassy
phenomena, the microscopic details being ``averaged out'' in this
case), hence the very strong increase of the relaxation time around
$T_{\rm g}$ should have a trace in spatial correlations.  One way to
address this puzzle is to define dynamical correlation lengths.
Convincing experiments~\shortcite{Bscience}, following numerical
simulations~\shortcite{Glotzer} and dynamical
theories~\shortcite{FPlength,BBmct} have, after a long search,
demonstrated for the first time growing dynamical
correlations~\shortcite{FPlength,BBmct,Bscience}. Another way, which
we shall review in this contribution, is to define a static
correlation length through a point-to-set procedure slightly more
involved than the two-point function underlying the definition of
the structure factor.

A large part of this chapter will be devoted to the discussion of
this length and its relation with the relaxation time.  In
Sec.~\ref{definition} we explain in details the definition of the
point-to-set correlation length and we show how it allows to prove
bounds between this correlation length and the equilibrium
correlation time that agrees with the intuition sketched above. We
then discuss two classes of disordered spin models of the 1RSB type
where spatial aspects can be addressed through analytic techniques,
namely models on diluted random graphs and finite dimensional models
in the Kac limit. In the first class of models, considered in
Sec.~\ref{corrRG}, each spin interacts with a finite number of other
spins chosen at random. These models can be solved exactly through
the cavity method; in agreement with the general bounds, at the
point of dynamical transition both the relaxation time and the
correlation length are divergent. The second class, studied in
Sec.~\ref{Kacmodels}, consists in genuine finite dimensional models
with a tunable interaction range $r_0$. In the limit of large $r_0$
one can compute the point-to-set correlation function and associated
correlation lengths. This leads to a detailed picture of glassy
phenomena, with a dynamical and static correlation lengths that do
not necessarily coincide.

We hope the style of presentation adopted in this chapter will
provide the reader with a global view and some mathematical and
theoretical  tools which should make easier the reading of the
original, more formal, literature on the subjects we address.

\section{Definition of the point-to-set correlation function and its relation to correlation time}
\label{definition}

\subsection{Heuristic discussion}
\label{sec_defpts_heuristic}

In this section we want to introduce the notion of point-to-set
correlation functions and to show that the correlation length
derived from it is relevant for glassy systems, as upper and lower
bounds on the correlation time can be inferred from this length. The
presentation will be first done in an informal way, following the
thought experiment first discussed in~\shortcite{BiBo}. We shall
then revisit it with more mathematical definitions and sketch the
results and the methods of proof of~\shortcite{MoSe}.

\begin{figure}
\centerline{\includegraphics[width=10cm]{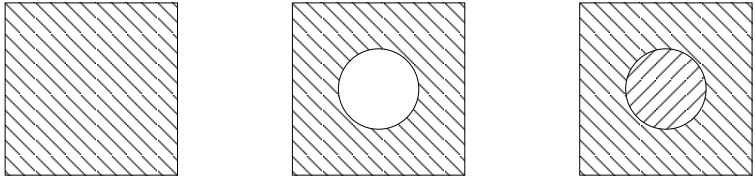}} \caption{Scheme of
the thought experiment~\protect\shortcite{BiBo} underlying the
definition of the point-to-set correlation length.} \label{fig_bb}
\end{figure}

Let us consider the thought experiment of~\shortcite{BiBo},
schematized in Fig.~\ref{fig_bb}. In a first time one takes a
snapshot of an equilibrium configuration of the system under study,
i.e. for instance the values of the spins, or the positions of
particles depending on the model investigated. Let us call $\us$
this first configuration, depicted on the first panel of
Fig.~\ref{fig_bb}. Suppose now that the configuration of the system
is frozen to the value it has in $\us$ outside a given volume
$\Bbar$ around an arbitrary point, for instance the center of the
system (see middle panel in Fig.~\ref{fig_bb}), and that the
interior is thermalized in presence of this boundary condition. One
thus obtains another equilibrium configuration $\us'$, which is
forced to coincide with $\us$ outside $\Bbar$. Consider now the
following question: how similar are $\us$ and $\us'$ around the
center of the system? A precise notion of similarity shall be given
below, in any case it is natural to expect that the larger the
volume $\Bbar$, the less similar should $\us$ and $\us'$ be at its
center. Indeed the influence of the boundary conditions, which force
$\us'$ to be very close to $\us$ when $\Bbar$ is small, becomes less
and less efficient when the boundary is pushed away. This procedure
thus allow to define a correlation function between a point (the
center of the system) and a set of points (the boundary of $\Bbar$),
hence the name already mentioned. It is understood that in the
correlation function the similarity measure should be averaged with
respect to the configurations $\us$ and $\us'$. From this function
one can further define a correlation length. Taking for simplicity
$\Bbar$ to be a spherical ball of radius $\l$, we shall indeed
define the correlation length $\lc$ as the minimal radius which
brings the point-to-set correlation function (i.e. the average
measure of similarity of the center of $\us$ and $\us'$) below a
small threshold fixed beforehand.

The equilibrium correlation time $\tc$ of the system can be defined
in a similar fashion, as the minimal time necessary for the
auto-correlation function (average similarity measure at the same
point, between one equilibrium configuration and the outcome of its
evolution during a certain amount of time) to drop below a given
threshold. It turns out that the intuition discussed in the
introduction, namely that large correlation times and large
correlation lengths are two intertwined phenomena, can be given a
precise content with these two definitions of $\tc$ and $\lc$.
Indeed the rigorous proof of~\shortcite{MoSe} we shall sketch below
implies that $\lc \le \tc \le \exp\left\{ \lc^d\right\}$, where we
have hidden for simplicity several constants. The interpretation of
these two inequalities might sound disappointingly simple. As
$\ell_{\rm c}$ measures the radius of a correlated region of the
system, and as for the center of the system to decorrelate it must
receive some information from the boundary of the correlated region,
the lower bound merely states that this information cannot propagate
faster than ballistically. On the other hand the upper bound follows
from the fact that the dynamics of the center of the system is
weakly sensitive to the outside of the correlated zone, hence it
should closely resemble the dynamics of the ball of radius $\lc$
without its surrounding environment. The latter case being the
dynamics of a finite system of volume $\lc^d$ (where $d$ denotes the
dimension of the system), its relaxation cannot be slower than
exponential in its volume.

The fact that these two inequalities have natural interpretations
does not mean they have a trivial content, but rather demonstrates
the relevance of the point-to-set definition of the correlation
length $\ell_{\rm c}$. Recall indeed that one of the puzzle of the
glass phenomenology is the drastic growth of the relaxation time
without significant traces in the structure factor, hence in the
2-point correlation length. The bounds between $\tc$ and $\lc$ show
that despite this fact the growth of the relaxation time must be
accompanied by a growth of a (static) correlation length if the
latter is appropriately defined.

\subsection{More precise definition of the correlation function}
\label{sec_defpts_precise}

We want now to provide the reader with more formal definitions of
the quantities discussed informally above, before restating with
more details the bounds between the correlation length and the
correlation time. Note that if point-to-set correlations are
relatively new in physics they are quite common in the mathematical
literature, in particular in the context of the so-called tree
reconstruction problem, see for
instance~\shortcite{MaSiWe,survey_reconstruction}.

\begin{figure}
%\vspace{-1cm}
\centerline{
\includegraphics[width=5cm]{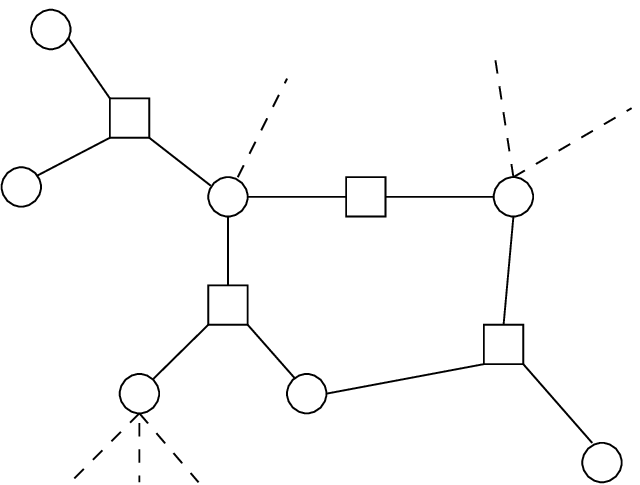}
\hspace{16mm} %\vspace{1cm}
\includegraphics[width=5cm]{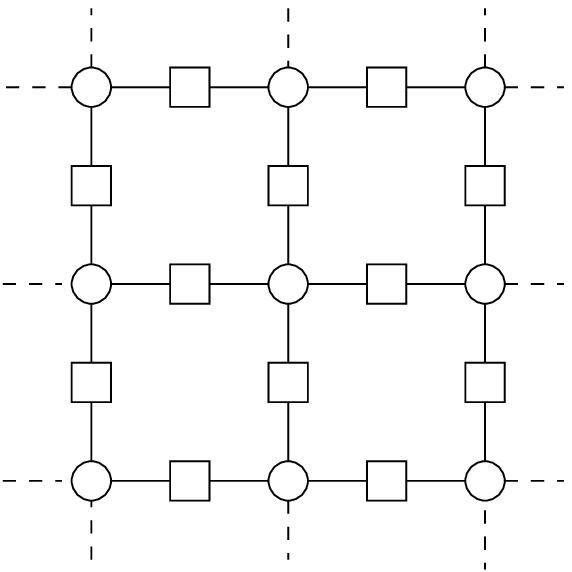}}
\caption{Left: general example of a factor graph representing an
energy $E(\us) = \sum_{a=1}^M E_a(\us_\da) $. Each circle represents
a variable $\s_i$, each square an interaction term $E_a$. An edge is
drawn between $E_a$ and $\s_i$ whenever the $a$-th interaction
depends on the $i$-th variable. Right: the case of a square lattice
with nearest neighbor interactions.} \label{fig_factor}
\end{figure}

For the sake of concreteness we shall consider a model of $N$ Ising
spins $\s_i =\pm 1$, whose global configuration will be denoted
$\us=(\s_1,\dots,\s_N)$. For a subset $S$ of the variable indexes
$\{1,\dots,N\}$ we will call $\us_S$ the configuration of the
variables in $S$. The energy function (Hamiltonian) is decomposed as
$E(\us) = \sum_{a=1}^M E_a(\us_\da) $, that is a sum of $M$ terms
$E_a$, with the $a$-th term involving a subset denoted $\da$ of the
variables. For instance in a two dimensional square lattice model
with nearest neighbor interactions there would be an interaction $a$
for each edge of the lattice. The definition of $E$ encompasses
however more general cases, in particular multi-spin interactions
involving more than a pair of spins. It can be convenient in such a
case to represent the network of interactions as a so-called factor
graph~\shortcite{fgraphs}, see Figure~\ref{fig_factor} for an
illustration. Each variable $\s_1,\dots,\s_N$ is associated to a
circle vertex, while the interactions $E_1,\dots,E_M$ are symbolized
with square vertices. An edge is drawn between a variable $i$ and an
interaction $a$ if and only if $E_a$ depends on $\s_i$.
%SF, in other words if and only if $i \in \da$.
On the right panel of Fig.~\ref{fig_factor} is drawn a portion of
the factor graph corresponding to a square lattice model.

We will use in the following the notation $d(i,j)$ for the distance
between two variables. This will be taken as the graphical distance
on the factor graph, that is the
%SF minimal
number of interactions that have to be crossed along a shortest path
%of the factor graph
linking $i$ and $j$. This notion of distance has the virtue of being
well-defined for any topology of the interaction network, not only
for finite-dimensional models. In the latter case the graph distance
is equivalent to the Euclidean one.

The Gibbs-Boltzmann probability measure is defined on the space of
configurations by $\mu(\us) = \exp[-\beta E(\us)]/Z$, with the
partition function $Z$ ensuring its normalization. Angular brackets
$\la \cdot \ra$ will be used to denote averages with respect to this
law. In order to quantify the amount of correlations induced by the
Gibbs-Boltzmann probability between one variable, say $\s_i$, and a
set $B$ of other variables, one can consider a function $F(\us_B)$
which depends only on the values of the variables in $B$, and
compute the correlation between the two as $\la \s_i F(\us_B) \ra -
\la \s_i \ra \la F(\us_B) \ra$. If for instance $\s_i$ were
completely independent from the status of the variables in $B$ this
quantity would vanish. To make contact with the thought experiment
explained above one can choose as a particular function $F$ the
magnetization of the $i$-th spin \emph{conditioned} on $\us_B$,
denoted $\la \s'_i \ra_{\us_B}$. This conditional average is defined
for an arbitrary function $f$ as
\begin{equation}
\la f(\us') \ra_{\us_B} = \sum_{\us'} f(\us') \mu(\us'|\us_B) \ ,
\end{equation}
where the conditional probability is restricted to configurations
which coincide with $\us_B$ on the variables in $B$,
\begin{equation}
\mu(\us'|\us_B) = \begin{cases} \frac{1}{Z(\us_B)} e^{-\beta
E(\us')} &
\text{if} \ \ \us'_B=\us_B \\
0 & \text{otherwise}
\end{cases} \ ,
\end{equation}
$Z(\us_B)$ ensuring its normalization. With this choice for $F$ we
have thus obtained the correlation function between a point $i$ and
a set $B$ as
\begin{equation}
G(i,B) = \la \s_i \la \s'_i \ra_{\us_B} \ra - \la \s_i \ra^2 =
\sum_{\us} \mu(\us) \sum_{\us'} \mu(\us'| \us_B) \s_i \s'_i - \la
\s_i \ra^2 \ , \label{eq_defG}
\end{equation}
where the last expression enlightens the connection with the
heuristic discussion above: the configuration $\us'$ corresponds to
the second equilibrium configuration drawn conditioned on the value
in $B$ of the first configuration $\us$.

We would like to emphasize that the configuration $\us'$ constructed
in the second place is as representative as $\us$ of the
Gibbs-Boltzmann equilibrium measure,  in other words for any
function $f$ of the configuration one has~\shortcite{elia}
\begin{eqnarray}
\la \la f(\us') \ra_{\us_B} \ra = \la f(\us) \ra. \label{obvious}
\end{eqnarray}
% This is probably obvious
% to the reader accustomed with conditional probabilities, for the sake of
% pedagogy we shall spell the derivation of this identity:
% \begin{eqnarray}
% \la \la f(\us') \ra_{\us_B} \ra &=&
% \sum_{\us} \mu(\us) \sum_{\us'} \mu(\us'| \us_B) f(\us')\\ &=&
% \sum_{\us} \frac{1}{Z} e^{-\beta E(\us)} \sum_{\us'} \frac{1}{Z(\us_B)}
% e^{-\beta E(\us')} \I(\us_B =\us_B') f(\us') \ ,
% \end{eqnarray}
% where here and in the following the indicator function $\I(X)$ is equal to
% 1 if $X$ is true, zero otherwise. We now denote $\Bbar$ the set of variables
% \emph{not} in $B$ and reorganize the sums as
% \begin{eqnarray}
% \la \la f(\us') \ra_{\us_B} \ra &=&
% \sum_{\us'} \frac{1}{Z} e^{-\beta E(\us')} f(\us')
% \sum_{\us_B,\us_\Bbar} \frac{1}{Z(\us_B)} e^{-\beta E(\us_B,\us_\Bbar)}
% \I(\us_B =\us_B') \nonumber \\
% &=& \sum_{\us'} \frac{1}{Z} e^{-\beta E(\us')} f(\us')
% \left[ \sum_{\us_\Bbar} \frac{1}{Z(\us'_B)} e^{-\beta E(\us'_B,\us_\Bbar)} \right] \ .
% \label{conditional}
% \end{eqnarray}
% One can finally notice that the term in square brackets is equal to one, and
% upon renaming $\us' \to \us$ one obtains the standard expression of
% $\la f(\us) \ra$.
%SF
This property %, which
follows simply from the properties of conditional probability and
its proof will be omitted. In the following sections we will
consider correlations between replicas of the system with the same
constraint (i.e. $\us'$ and $\us''$ are generated independently from
$\mu(\cdot|\us_B)$ for fixed $\us_B$ and quenched disorder in the
energy function $E$). Given any two functions $f(\us)$ and $g(\us)$
one can define the correlation
\begin{eqnarray}
C(f,g)=\langle \langle f(\us')\rangle_{\us_B} \langle
g(\us'')\rangle_{\us_B} \rangle \ .
\end{eqnarray}
Similarly to (\ref{obvious})
% Using the same reasoning as the one leading to (\ref{conditional})
one can show that this coincides with the correlation between $\us'$
and $\us$ according to
\begin{eqnarray}
C(f,g)=\langle f(\us) \langle g(\us')\rangle_{\us_B}\rangle.
\label{equal}
\end{eqnarray}
This identity will play an important role in the analysis of the
point-to-set function in the Kac limit discussed in
Sec.~\ref{Kacmodels}, and we shall call it conditional equilibrium
condition.

We come back now to the definition of the point-to-set correlation
function, and consider more specifically the case where $B$ is the
outside of a ball of radius $\ell$ around $i$, i.e. the set of
variables at distance larger or equal than $\ell$ from variable $i$,
$B(i,\ell)=\{j|d(i,j)\ge \ell\}$. We shall call
$G(i,\ell)=G(i,B(i,\ell))$ the correlation function for this
geometry. It is intuitively clear that $G(i,\ell)$ decreases when
the radius $\ell$ of the ball increases, as farther away sites are
less correlated with $i$. One can thus set a small threshold $\ve$
and defines the correlation length for site $i$ as the minimal
distance $\ell$ necessary to make the correlation $G(i,\ell)$ drops
below the threshold $\ve$. In formula, $\ell_i(\ve) = \min\{\ell |
G(i,\ell) \le \ve \}$.

We turn finally to the definition of the correlation time. We shall
consider a single spin flip dynamics in continuous time, defined
through transition rates $W(\us \to \us')$. The single spin flip
assumption means that these rates vanish whenever $\us$ and $\us'$
differ in more than one variable. We assume the rates to verify the
detailed balance condition
\begin{equation}
\mu(\us) W(\us \to \us') = \mu(\us') W(\us' \to \us) \ ,
\label{eq_bd}
\end{equation}
which ensures that the Gibbs-Boltzmann probability is stationary
under this dynamics. Moreover the rate from a configuration $\us$ to
the configuration where a single variable $i$ has been flipped is
assumed to depend on $\us$ only through the configuration of $\s_i$
and of the variables at unit distance from $i$. This last condition
is obviously fulfilled by the usual Monte Carlo dynamics like the
Metropolis or heat-bath (Glauber) rules. The equilibrium dynamics of
the model is defined by these rates and the initial condition
$\us(t=0)$ which is drawn from the equilibrium law $\mu$. The
average over the initial condition and the subsequent evolution
shall be denoted again by angular brackets $\la \dots \ra$. The
auto-correlation function of variable $i$ is hence defined as
$C_i(t) = \la \s_i(0) \s_i(t) \ra - \la \s_i \ra^2$, and we can
assign a correlation time $\t_i$ to this variable as $\t_i (\ve) =
\min \{ t | C_i(t) \le \ve  \} $, i.e. as the minimal time for the
auto-correlation function to drop below a given threshold $\ve$. To
avoid confusion in the following let us emphasize that in the
definition of the correlation time we consider the dynamics of the
\emph{whole} system; the constraints on the finite ball $B$ only
appear in the definition of the correlation length.

\subsection{Relation between the correlation length and the equilibrium
correlation time} \label{relcorrtime}

Now that the definitions of the correlation time $\t_i$ and
correlation length $\ell_i$ have been given more precisely we can
restate in a more accurate way the bounds derived rigorously
in~\shortcite{MoSe}:
\begin{equation}
C_1\, \ell_i(\ve') \le \tau_i(\ve) \le 1+\exp\left\{C_2
|\Bbar(i,\ell_i(\ve''))| \right\} \ , \label{eq_bound}
\end{equation}
where $|\Bbar|$ denotes the number of sites in the ball $\Bbar$. In
a finite dimensional setting one has $|\Bbar(\ell)|\sim \ell^d$, but
Eq.~(\ref{eq_bound}) is valid for any topology of the interaction
graph. The small thresholds $\ve'$ and $\ve''$ are functions of
$\ve$ which go to zero when $\ve$ vanishes, and $C_{1,2}$ are
numerical constants which depends on the microscopic details of the
Hamiltonian and of the dynamics.

Let us make a series of remarks on this result:
\begin{itemize}

\item For simplicity we have explained this relationship between length and
time scales for a discrete system of Ising spins, with arbitrary
interactions. Its extension to more general discrete degrees of
freedom is simple and was the case considered in~\shortcite{MoSe}.
It is natural to expect that a similar result will hold for particle
systems evolving in the continuum, for instance by means of a coarse
grained occupation number field, see e.g.~\shortcite{CaGrVe}.

\item The relationship between $\ell_i$ and $\t_i$ holds site by site. This is
particularly important for inhomogeneous systems with quenched
disorder, where the correlation lengths and times can vary wildly
from site to site.

\item In the limit of zero temperature the constant $C_2$ will diverge: the
presence of trivial ``energetic barriers'' can lead to very large
correlation times without a growing correlation length.

\item The lower and upper bound in Eq.~(\ref{eq_bound})
are widely separated when $\ell$ grows, which could suggest that the
bounds are very far from optimal. However it can be argued that with
the very weak hypotheses made for its derivation, this result can
only be marginally enhanced (i.e. at the level of the numerical
constants and with a dynamical exponent $z \ge 2$ in the lower
bound). Indeed both bounds can be saturated in the low/high
temperature regimes of some models which enters in the range of
validity of the result (see below for a discussion of diluted
mean-field models).

\item We would like to emphasize the \emph{static} character of the
point-to-set definition of the correlation length. Indeed the
expression of the correlation function stated in Eq.~(\ref{eq_defG})
involves only equilibrium averages, and does not make any reference
to the dynamical evolution of the system. Of course if one wants in
practice to evaluate this function for a system or a model which
does not admit an analytical solution, Monte Carlo simulations will
probably have to be used to generate thermalized configurations of
the full and constrained systems, yet this dynamics is here only a
computational tool and not a part of the definition.

\item The dependence of the correlation length and time on the arbitrary
threshold $\ve$ may look rather unsatisfactory at first sight.
Fortunately this is not a real issue for glassy systems: one expects
for them a discontinuous behaviour upon approaching the glass
transition. The spatial and temporal correlation function decay to
zero in a two-step fashion, with the appearance of a growing
plateau. As long as $\ve$ is smaller than the height of the plateau
(called Edwards-Anderson or non-ergodicity parameter) the asymptotic
behaviour of $\t(\ve)$ and $\ell(\ve)$ is essentially independent on
the choice of this threshold. This is however a concern for more
conventional critical phenomena where the order parameter grows
continuously at the transition.

\item It should be acknowledged that these bounds do not apply directly for
kinetically constrained models (KCM)~\shortcite{FA,KA}. As a matter
of fact the equilibrium measure of these models is a trivial product
measure factorized over the sites, hence the correlation length as
defined above is always equal to 1. However the dynamical rules
defining the KCM violates a technical assumption of permissivity
which is necessary for the derivation of these bounds.

\end{itemize}

\subsection{Determination of the correlation length in finite dimensional
systems} In the following subsection we shall give some explanations
on how these bounds can be proven, then the focus of the rest of the
chapter will be put on two families of models where analytical
computations can be pushed further. Before that we want however to
mention some papers, mostly numerical, where finite dimensional
models have been investigated with a perspective somehow related to
the point-to-set correlation function.

To the best of our knowledge the first papers where glassy systems
in confined geometry with  boundary conditions self-consistently
generated according to the prescription described in the previous
sub-sections were in ref.~\shortcite{KobPS}. These works
investigated the effect of the presence of a wall or confining
geometries on the dynamics of Lennard-Jones binary mixtures.

More recent literature has focused on the thought experiment
of~\shortcite{BiBo}. In~\shortcite{Rob} a plaquette spin model of a
glass was studied analytically and numerically. The effect of
boundary conditions on finite size subsystems could be analyzed and
led to a determination of a static correlation length along the
lines of the thought experiment of~\shortcite{BiBo}.

\begin{figure}
\includegraphics[width=10cm]{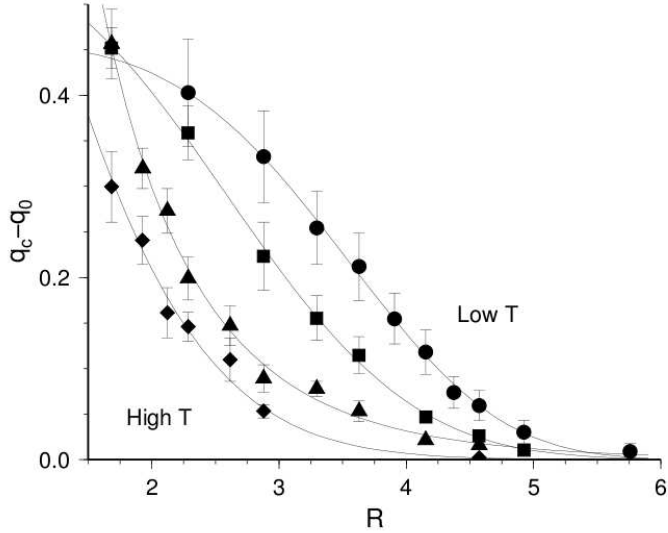}
\caption{Point-to-set correlation function for a binary mixture of
soft-sphere particles, from~\protect\shortcite{BiBoCaGrVe}. When the
temperature is lowered the influence of the boundary of a spherical
cavity persists for larger radius $R$.} \label{fig_BiBoCaGrVe}
\end{figure}

A binary mixture of soft-sphere particles (i.e. a fragile
glass-former liquid modeled microscopically with particles in the
continuum) was considered
in~\shortcite{CaGrVe,BiBoCaGrVe,CaCaGrGrVe}. The procedure described
above of freezing an equilibrium configuration outside a cavity of a
given radius and thermalizing its interior was thus implemented with
Monte Carlo simulations. This led to a demonstration of the growth
of the static correlation length upon lowering the temperature of
the liquid. In Fig.~\ref{fig_BiBoCaGrVe} we reproduce a result
of~\shortcite{BiBoCaGrVe}, which shows the point-to-set correlation
as a function of the radius of the cavity, for various temperatures.
% A related study in a slightly different geometric configuration
%can be found in~\shortcite{Kob}.
The reader is also referred
to~\shortcite{CaGrVe,BiBoCaGrVe,CaCaGrGrVe} for the interpretation
of these results along the lines of RFOT and for a critical
discussion of the scaling exponents of RFOT.

\subsection{Tools for the proof}

\subsubsection{Couplings}

In this subsection we shall sketch in an informal way the method of
proof of Eq.~(\ref{eq_bound}), the reader being referred
to~\shortcite{MoSe} for the details. We want in particular to
introduce a very useful probabilistic concept on which the proof of
both the upper and the lower bounds of~(\ref{eq_bound}) relies,
namely the construction of a coupling between stochastic
processes~\shortcite{coupling}. Let us first explain what is a
coupling on the simplest case of two random variables, for instance
two biased coins. The first coin $X^{(1)}$ takes value $H$ead with
probability $p_1$, $T$ail otherwise, while the probability of $H$ead
of the coin $X^{(2)}$ is $p_2$. A coupling of these two biased coins
is a random variable $(X^{(1)},X^{(2)})$ which can take four values
$\{(H,H),(H,T),(T,H),(T,T)\}$, such that if one observes only the
first (resp. second) element of the couple $(X^{(1)},X^{(2)})$, one
sees $H$ead occurring with probability $p_1$ (resp. $p_2$).
Obviously one trivial way to construct such a coupling is to take
$X^{(1)}$ and $X^{(2)}$ as independent copies of the original biased
coins. The power of the notion of coupling relies in the possibility
to introduce a dependency between the two elements of the couple,
without spoiling the partial (marginal) frequency of observations of
head in the first or second position. It is indeed easy to realize
that there exists an infinity of couplings of these two biased coins
parametrized by a real number. Let us give one explicit example.
Suppose without loss of generality that $p_1 \ge p_2$, and set the
value of the coupling to be
\begin{equation}
(X^{(1)},X^{(2)}) = \begin{cases}
(H,H) & \text{with probability} \ \ p_2 \ , \\
(H,T) & \text{with probability} \ \ p_1-p_2 \ , \\
(T,T) & \text{with probability} \ \ 1-p_1 \ .
\end{cases}
\label{pdilu}
\end{equation}
One can easily check that this is indeed a coupling (the marginal
probabilities for the $H$ead of the two coins are respectively $p_1$
and $p_2$), and that, for given values of $p_1$ and $p_2$, it
minimizes the probability that $X^{(1)} \neq X^{(2)}$. For this
reason it is termed the \emph{greedy coupling} of the two random
variables. In particular if the two coins are identical, $p_1=p_2$,
then $X^{(1)}$ is always equal to $X^{(2)}$ in this coupling.

\subsubsection{Lower bound}

This notion of coupling extends naturally to random variables more
complicated than biased coins, and also to stochastic processes,
that is sequences of random variables indexed by a time parameter.
The proof of the lower bound in Eq.~(\ref{eq_bound}) relies indeed
on such a construction that we shall now explain.

Let us choose a variable $i$ and a positive integer $\ell$, and
denote as above $B$ (resp. $\Bbar$) the outside (resp. inside) of
the ball of radius $\ell$ around $i$. Consider now the stochastic
process $(\us^{(1)}(t),\us^{(2)}(t))$, where two configurations of
the same system evolves simultaneously, defined by the following
rules:
\begin{itemize}
\item at the initial time $t=0$, the two configurations coincide,
$\us^{(1)}(0)=\us^{(2)}(0)=\us$, with $\us$ drawn from the
equilibrium Gibbs-Boltzmann measure.
\item at any time $t>0$ where a variable $j$ attempts an update of its value,
both configurations are modified together according to:
\begin{itemize}
\item if $j\in B$, the configuration $\us^{(2)}(t)$ is kept unchanged, while
the spin $\s_j^{(1)}(t)$ is flipped or not according to the
transition rates $W(\us^{(1)}(t) \to \us')$.
\item if $j \in \Bbar$, one determines the probability of $\s_j^{(1)}(t)$
(resp. $\s_j^{(2)}(t)$) right after the update according to
$W(\us^{(1)}(t)\to \us')$ (resp. $W(\us^{(2)}(t) \to \us')$). Then
the new values of $(\s_j^{(1)}(t),\s_j^{(2)}(t))$ are drawn
according to the greedy coupling (as defined above) of these two
Ising spin random variables.
\end{itemize}
\end{itemize}
Observed separately each element of this coupling corresponds :
\begin{itemize}
\item for
$\us^{(1)}(t)$ to the original dynamics of the whole system.
\item for
$\us^{(2)}(t)$ to the equilibrium dynamics for the inside of the
ball $\Bbar$, submitted to a time-independent boundary condition in
$B$, with $\us_B^{(2)}(t)$ fixed forever to the value $\us_B$ it has
at time $t=0$.
\end{itemize}
It turns out that the original temporal and point-to-set correlation
functions can be computed from appropriate averages over this
coupling, that we shall still denote $\la \dots \ra$ with a slight
abuse of notation. Assuming that the equilibrium magnetization $\la
\s_i \ra$ vanishes to simplify the discussion, one realizes that
$C_i(t) = \la \s_i^{(1)}(0) \s_i^{(1)}(t)\ra$, while $G(i,B)$ is the
limit of $\la \s_i^{(2)}(0) \s_i^{(2)}(t)\ra$ as $t$ goes to
infinity. To obtain the lower bound of Eq.~(\ref{eq_bound}) one has
to relate in some way the behavior of the spatial and temporal
correlation functions. From the above observation this translates
into a comparison of the value of $\s_i$ in the two processes
$\us^{(1)}(t)$ and $\us^{(2)}(t)$. The idea of the proof is then to
exploit the properties of the above defined coupling. In fact for
``small'' times, that is of order smaller than $\ell$, most probably
$\s_i^{(1)}(t)=\s_i^{(2)}(t)$. Indeed, at initial times the two
configurations coincide everywhere in the system. Moreover, when
$j\in\Bbar$ tries to update, if the two configurations coincide on
$j$ and its immediate neighborhood, then necessarily
$\s_j^{(1)}=\s_j^{(2)}$ right after the update, because of the use
of the greedy coupling of the two update probabilities. In other
words, disagreement between the two configurations can only take
birth in the outside $B$ of the ball of radius $\ell$, and must
propagate from the surface of the ball to its
center~\shortcite{disag1,disag2}.
%SF
% , a phenomenon called
%\emph{disagreement percolation}

Let us rephrase this reasoning with more explicit formulas. We start
from a simple inequality on the point-to-set correlation function,
\begin{equation}
G(i,B) = \lim_{t \to \infty} \la \s_i^{(2)}(0) \s_i^{(2)}(t)\ra \le
\la \s_i^{(2)}(0) \s_i^{(2)}(t)\ra \ , \label{eq_decreasing}
\end{equation}
which holds for any value of $t\ge 0$. Indeed the equilibrium
auto-correlation functions of reversible Markov processes are
decreasing in time\footnote{This is a simple property that can be
proven using the spectral decomposition of the evolution operator
$W(\s\to\s')$.}. To relate the two coupled processes let us define
the indicator function $X(t)=\delta_{\s_i^{(1)}(t),\s_i^{(2)}(t)}$.
We can thus upper-bound the spatial correlation function as
\begin{eqnarray}
%SF
G(i,B) & \le & \la \s_i^{(1)}(0) \s_i^{(1)}(t) \ra-\la
\s_i^{(1)}(0) \s_i^{(1)}(t)(1- X(t))\ra + \la \s_i^{(2)}(0)
\s_i^{(2)}(t) (1-X(t))\ra \nonumber
 \\
& \le & \la \s_i^{(1)}(0) \s_i^{(1)}(t) \ra + 2 \la (1-X(t)) \ra
\label{eq_lb_last_m1} \nonumber
\\
& = & C_i(t) + 2 \ {\mathcal P}_{\rm dis}(t) \ , \label{eq_lb_last}
\end{eqnarray}
where ${\mathcal P}_{\rm dis}(t)$ is the probability that the two
parts of the coupling disagrees on the value of $\s_i$ at time $t$.
% ~\footnote{The
% suspicious reader might wonder about the origin of the factor 2 in
% (\ref{eq_lb_last_m1}). This is because one is not allowed to write
% $\la \s_i^{(1)}(0) \s_i^{(1)}(t) X(t)\ra \le
% \la \s_i^{(1)}(0) \s_i^{(1)}(t) \ra$, as $\s_i^{(1)}(0) \s_i^{(1)}(t)$
% can be negative. Instead one has to expand
% $\la \s_i^{(1)}(0) \s_i^{(1)}(t) X(t)\ra =
% \la \s_i^{(1)}(0) \s_i^{(1)}(t) (1-(1-X(t))\ra \le
% \la \s_i^{(1)}(0) \s_i^{(1)}(t) \ra + {\mathcal P}_{\rm dis}(t)$.}.
As explained above disagreement between the two configurations of
the coupling has to travel from the boundary of $B$ towards $i$
along a a sequence of $\ell$ adjacent spins on a path from $B$ to
$i$. An upper bound on ${\mathcal P}_{\rm dis}(t)$ can thus be
obtained by multiplying the number of such paths with the
probability that in the interval of time $[0,t]$ all $\ell$ sites of
a given path attempts to update their configurations in the right
order (from the boundary inwards). The first factor is obviously
smaller than the maximum connectivity of a spin raised to the power
$\ell$. Because the times where the spin attempts an update form a
Poisson process the second factor is the probability than a Poisson
random variable of average $t$ is greater than $\ell$, and this last
probability is smaller than $(e \, t /\ell)^\ell$. Putting these two
factors together one obtains that ${\mathcal P}_{\rm dis}(t) \le (C
t/\ell)^\ell$, where $C$ is a constant which depends on the
connectivity of the interaction graph. Now if one sets
$t=\tau_i(\ve)$, the r.h.s. of (\ref{eq_lb_last}) can be made
smaller than $2 \ve$ by taking $\ell$ larger than some constant
multiplied by $\tau_i(\ve)$, hence the lower bound in
Eq.~(\ref{eq_bound}).

\subsubsection{Upper bound}

The upper bound of Eq.~(\ref{eq_bound}) is obtained by showing that
the autocorrelation function $C_i(t)$ of the $i$-th spin is weakly
sensitive to the configuration of the system out of the ball of
radius $\ell_i(\ve)$. One can thus approximate $C_i(t)$ with the
value it would have in the system made only of the interior $\Bbar$
of the ball of radius $\ell_i(\ve)$. It is natural that the
autocorrelation time of the latter cannot grow faster than
exponentially in its volume. Consider indeed an arbitrary system
with $n$ variables, and the following coupling between two copies of
its dynamics. It is initialized with two arbitrary configurations
$(\us^{(1)}(0),\us^{(2)}(0))$, and one performs at later times the
updates of $(\s_j^{(1)}(t),\s_j^{(2)}(t))$ according to the greedy
coupling of the two transition rates. This implies that as soon as
the two copies coincide, they remain the same for all subsequent
times. Moreover we assume the dynamics to be ``permissive'', that is
there is a constant $\kappa > 0$ such that when $j$ attempts an
update, irrespectively of the neighborhood of $j$ in
$(\us^{(1)},\us^{(2)})$, the probability that
$\s_j^{(1)}=\s_j^{(2)}$ after the update is larger than $\kappa$.
This condition of permissivity is satisfied at any strictly positive
temperature if there are no hard constraints in the model (no
infinite energy configurations), with possibly $\kappa$ vanishing
when $T \to 0$. We can upper-bound the relaxation time by the
coalescence time of the coupling: if for two arbitrary initial
configurations (as different as they can be) the coupled dynamics
has coalesced at a given time $t_0$, hence ``forgotten'' its initial
conditions, then the equilibrium dynamics of a single configuration
would have as well. What remains to be proven is that, with a large
probability, this coalescence time is not larger than exponential in
the number of variables $n$. This follows from considering a
particular sequence of update events that brings the two evolving
copies of the system to coincide, namely that all variables have
attempted to update at least once, and that their last update
brought the variables to the same value in the two parts of the
coupling. The probability of the last condition is by definition
larger than $\kappa^n$, hence on time intervals larger than
$\kappa^{-n}$ coalescence is very probable.

\section{Computation of the correlation function in mean-field (random graph) models}
\label{corrRG}

In this and the next section we shall discuss simplified models, of
mean-field nature (in a sense that shall be precised), for which
analytical computations of the point-to-set correlation function are
possible.

\subsection{A reminder on mean-field glassy models}

Many researchers agree that important insights on the physics of
glassy systems have been gained by the study of apparently remote
models, namely mean-field spin glasses with multi-spin interactions.
The paradigmatic example is the so-called $p$-spin model, with $p
\ge 3$, defined by the Hamiltonian
\begin{equation}
E(\us)=-\underset{1\le i_1<\dots <i_p\le N}{\sum} J_{i_1\dots i_p}
\s_{i_1}\dots \s_{i_p} \ , \label{eq_def_pspin}
\end{equation}
where the $J_{i_1\dots i_p}$ are independent Gaussian random
variables of zero mean and variance $\mathbb{E}[J_{i_1\dots
i_p}^2]=\frac{p!}{2 N^{p-1}}$. In the Ising version of the model the
degrees of freedom are $\s_i = \pm 1$, while the $\s_i$ are reals in
the spherical case, with the additional constraint $\sum_i \s_i^2 =
N$. The properties of this model have been extensively studied, we
refer the reader to the lecture
notes~\shortcite{Houches_Giorgio,Houches_Leticia,lect_pspin} and
references therein for the references to the original works. We
shall content ourselves here with a very brief reminder of its
salient features. From a thermodynamical point of view, the
free-energy of this model exhibits a non-analytic behaviour at the
critical temperature $T_{\rm c}$ (also called Kauzmann temperature).
However the dynamical properties of the model changes at the dynamic
transition temperature $T_{\rm d} > T_{\rm c}$: the Gibbs measure
gets split onto an exponential number of pure states, the
equilibration time of the system diverges in the thermodynamic limit
for $T < T_{\rm d}$, hence the dynamics exhibit aging (with a
non-trivial effective temperature) if the system is initialized in a
random configuration. The absence of any thermodynamic singularity
at $T_{\rm d}$ is due to a compensation phenomenon between the
exponential degeneracy of the pure states (complexity, or
configurational entropy) and the internal free-energy of the pure
states, such that the total free-energy of the system in the
intermediate phase between $T_{\rm d}$ and $T_{\rm c}$ coincides
with the analytical continuation of the high temperature (liquid)
phase. In technical words one has a one step of replica symmetry
breaking (1RSB) behaviour with Parisi breaking parameter $m=1$. On
the contrary at $T_{\rm c}$ the complexity vanishes, and for
$T<T_{\rm c}$ the system enters a true 1RSB phase with $m<1$.

Of course the main criticism which can be made against the relevance
of such a model for the study of structural glasses is its
completely different microscopic nature, in particular because of
the quenched disorder present in its definition. The two variations
of the $p$-spin model presented in this and in the next section
shall not escape this criticism, yet they partially cure some of the
pathologies of the model and present an even richer phenomenology,
which motivates the theoretical interest for them.

\subsection{Definition of diluted mean-field models}

The model defined by the Hamiltonian (\ref{eq_def_pspin}) together
with the above stated variance of the couplings constants
$J_{i_1\dots i_p}$ will be termed the fully-connected $p$-spin
model. Indeed all of the ${N \choose p}$ possible interactions
between $p$-uplets of spins are present in the system, even if each
of them is individually weak in order to have an extensive scaling
of the Hamiltonian in the thermodynamic limit. In this case each
spin interacts with all the others, in other words the distance
$d(i,j)$ between any pair of spins defined in
Sec.~\ref{sec_defpts_precise} is equal to one. This peculiarity,
very different from the finite-dimensional models with short-range
interactions where any degree of freedom interacts only with a
finite number of neighbors, can be cured with the following
definition of the coupling constants:
\begin{equation}
J_{i_1\dots i_p} = \begin{cases}
%SF
\pm 1 & \text{with probability} \ \alpha\frac{p!}{2 N^p} \\
%-1 & \text{with probability} \  \alpha \frac{p!}{2 N^p} \\
0 & \text{with probability} \ 1 - \alpha \frac{p!}{N^p}
\end{cases} \ .
\label{eq_def_Jdiluted}
\end{equation}
This case will be called the diluted $p$-spin model. At variance
with the fully-connected case, only an extensive number
(concentrated around its average $\alpha N$) of interactions are now
present, but each of them is strong (of order 1). Moreover a short
computation reveals that the degree of one spin, that is the number
of interactions it belongs to, converges in the thermodynamic limit
to a Poisson random variable of average $\alpha p$, hence finite
with respect to the system size $N$.

This kind of diluted model, pioneered in the case $p=2$ by Viana and
Bray~\shortcite{ViBr}, can be viewed as intermediary between
fully-connected and finite-dimensional ones. They share with the
latter the finiteness of the connectivity of each degree of freedom,
but are mean-field as the former ones. Indeed there is no a priori
underlying Euclidean space in their definition. This allows for
detailed analytical computation of their properties, and provides a
family of models where the Bethe-Peierls approximation is actually
exact in the thermodynamic limit. Moreover there is one more control
parameter besides the temperature, namely the ``density'' parameter
$\alpha$ which controls the number of interactions present in the
system.

There has been an important research effort in the last decades in
the community of statistical mechanics of disordered systems to
develop analytical tools for tackling such diluted spin-glasses
models. An important motivation has been the intimate connection
between these models and random combinatorial optimization problems,
as the $k$-satisfiability and the coloring of random graphs. An
extensive account of this line of research can be found in the
recent book~\shortcite{MeMo_book}.

\begin{figure}
\centerline{\includegraphics[width=8cm]{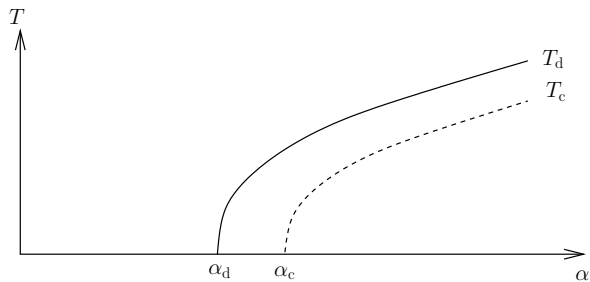}}
\caption{Shape of the phase diagram for the diluted $p$-spin model.}
\label{fig_plot_sketch_dilpspin}
\end{figure}

Before turning to some explanations on the computations in diluted
models, in particular the determination of the point-to-set
correlation function, let us plot in
Fig.~\ref{fig_plot_sketch_dilpspin} the shape of the phase diagram
in the (density $\alpha$, temperature $T$) plane. The two transition
temperatures $T_{\rm d}$ and $T_{\rm c}$ of the fully-connected
model (which corresponds to the $\alpha \to \infty$ limit up to a
rescaling of the energy) becomes two lines of transitions, which
ends up at zero temperature for finite values of the connectivity
$\alpha$, respectively $\alpha_{\rm d}$ and $\alpha_{\rm c}$. The
zero-temperature properties of this model have been largely studied
in the context of optimization problems (where it is known as
XORSAT), the thresholds $\alpha_{\rm d}$ and $\alpha_{\rm c}$
corresponding there to the clustering and satisfiability
transitions~\shortcite{xor1,xor2}.

\subsection{Computation of the correlation function}
\label{sec_dil_computation}

We shall now explain the main steps in the computation of the
point-to-set correlation function in mean-field diluted systems. Let
us first discuss the properties of the (random) factor graph
associated with the diluted $p$-spin model defined by
Eqs.~(\ref{eq_def_pspin},\ref{eq_def_Jdiluted}). To make contact
with the general definitions we rewrite it as
\begin{equation}
E(\us) = \sum_a E_a(\us_\da) \ , \qquad E_a(\us_{\da}) = - J_a
\prod_{i \in \da} \s_i \ , \label{eq_def_pspin2}
\end{equation}
where we have only retained the $p$-uplets with a non-vanishing
coupling constant and renumbered them in terms of the interaction
index $a$. Each interaction $a$ involves $p$ variables, while each
variable appears in a random number of interactions, which is easily
found to be a Poisson random variable of average $\alpha p$. What is
slightly less obvious is the fact that these random factor graphs
have a local tree structure, that is short loops are relatively
rare, and only loops of length of order $\log N$ begin to
proliferate in the thermodynamic limit ($N \to \infty$ with $\alpha$
finite). It is this locally tree-like property which allows for
analytical computations in these models; note that this property is
shared by other definitions of the random factor graph model, in
particular with degree distributions other than Poissonian, for
instance deterministic.

\begin{figure}
\centerline{\includegraphics[width=8cm]{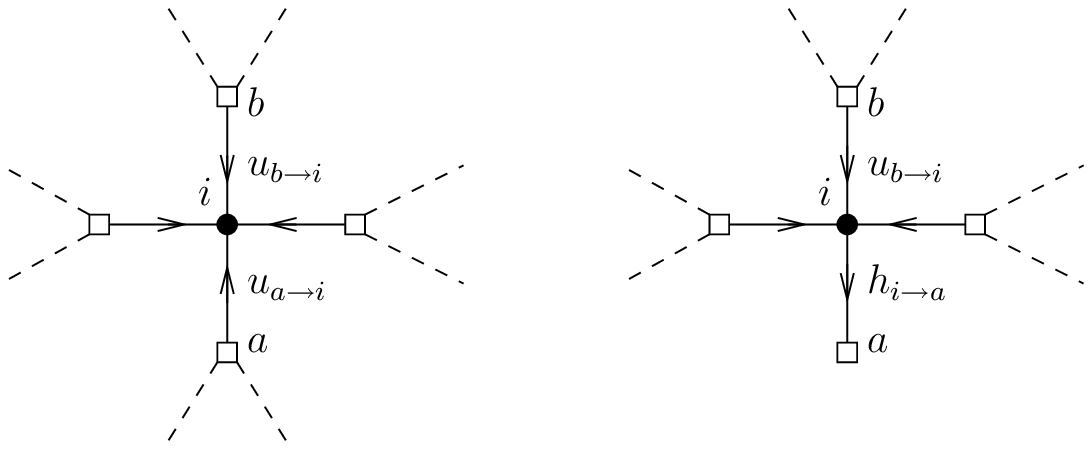}}
\caption{Pictorial representation of the recursive computation on a
tree. Left: the effective magnetic field $H_i$ is obtained from the
effective fields in the sub-trees $F_{a \to i}$, see
Eq.~(\ref{eq_recurs_H}). Right: the construction of the field for
$F_{i \to a}$ from the $F_{b \to i}$, as shown in the second part of
Eq.~(\ref{eq_recurs_hu}).} \label{fig_eq_recurs}
\end{figure}

Let us forget for a moment the existence of (long) loops in the
factor graph and see how the problem can be solved easily on a
finite tree. The key point is the natural recursive structure of a
tree: removing a vertex from a tree leads to a set of smaller
sub-trees, which, by definition, are disconnected one from the
others. So if one can solve the considered problem on trees of some
given size, and glue together the solutions to compute the solution
on the concatenated tree, then by recurrence the solution can be
found for any tree. Consider for instance the computation of the
local magnetization $\la \s_i \ra$ of a spin in a tree factor graph.
According to the recursive strategy just explained, this can be
expressed in terms of the magnetization in the sub-trees $F_{a \to
i}$ where only one interaction $a$ around $i$ is retained, in
formula one defines the effective magnetic field $H_i$ acting on
$i$,
\begin{equation}
\la \s_i \ra = \tanh(\beta H_i) \ ,
\end{equation}
and deduces the latter from the effective fields in the sub-trees
$F_{a \to i}$,
\begin{equation}
H_i = \sum_{a \in \di} u_{a \to i} \ . \label{eq_recurs_H}
\end{equation}
This relation is represented schematically on the left panel of
Fig.~\ref{fig_eq_recurs}.
These in turns can be computed from the effective fields in the
sub-trees of the form $F_{j \to a}$, in which all interactions
around the variable $j$ except $a$ is retained,
\begin{equation}
u_{a \to i} = f_{a \to i}(\{h_{j \to a}\}_{j \in \dami}) \ , \qquad
h_{i \to a} =\sum_{b \in \dima} u_{b \to i} \ , \label{eq_recurs_hu}
\end{equation}
see the right panel of Fig.~\ref{fig_eq_recurs} for an illustration.
The function $f_{a \to i}$ depends on the precise form of the
interaction energy $E_a$; in the case of the $p$-spin interaction
defined in Eq.~(\ref{eq_def_pspin2}) one finds
\begin{equation}
u_{a \to i} = \frac{1}{\beta} {\rm arctanh} \left[\tanh(\beta J_a)
\prod_{j \in \dami} \tanh(\beta h_{j \to a}) \right] \ ;
\end{equation}
the present discussion applies to any kind of interactions $E_a$,
provided one takes the function $f_{a \to i}$ corresponding to the
model under study.

To summarize, the equilibrium properties of a tree factor graph
model can be obtained by solving the recursion equations
(\ref{eq_recurs_hu}) for the effective fields $\{h_{i \to a},u_{a
\to i}\}$ living on the directed edges of the factor graph. These
equations admit a unique solution which can be easily obtained by
starting from the leaves of the tree (the variables with a unique
neighboring interaction), where the boundary condition is $h_{i \to
a}=0$.

We can now proceed with the computations of the conditional averages
$\la \dots \ra_{\us_B}$, where the configurations of the variables
in a subset $B$ of the variables is fixed to $\us_B$. A moment of
thought reveals that this is a simple modification of the previous
computation. In fact one has to look for a set of effective fields
$\{h_{i \to a}^{\us_B},u_{a \to i}^{\us_B}\}$, which depends on the
imposed configuration $\us_B$. They obey the same recursive
equations (\ref{eq_recurs_hu}), but with an additional boundary
condition: $h^{\us_B}_{i \to a}=\s_i \times \infty$ if $i \in B$.
Indeed a fixed variable feels an infinite effective field (positive
or negative according to the value it is fixed to). Once the fields
$\{h_{i \to a}^{\us_B},u_{a \to i}^{\us_B}\}$ have been determined
on all edges of the factor graph one obtains in a simple way the
conditional magnetizations, for instance
\begin{equation}
\la \s'_i \ra_{\us_B} = \tanh(\beta H_i^{\us_B}) \ , \qquad
H_i^{\us_B} = \sum_{a\in \di} u_{a \to i}^{\us_B} \ .
\end{equation}

To complete the computation of the correlation function defined in
equation (\ref{eq_defG}), one has to average the correlation $\s_i
\la \s'_i \ra_{\us_B}$ with respect to $\s_i$ and $\us_B$, which are
both parts of a reference equilibrium configuration $\us$. The
generation of an equilibrium configuration on a tree factor graph is
most naturally done in a recursive, broadcasting like, way. This
means that one has first to choose the value of a reference
variable, say $i$, according to its marginal probability law. The
latter is known on a tree, since the probability of $\s_i$ to be
$+1$ in an equilibrium configuration is by definition $e^{\beta
H_i}/(2 \cosh(\beta H_i))=(1+\la \s_i \ra)/2$. Once the value of
$\s_i$ is fixed in this way, the problem of the generation of the
other variables decouples between the various branches emanating
from $i$, thanks again to the tree structure of the factor graph.
One can treat independently each interaction $a$ around $i$, and
generate the value of the other variables appearing in $a$,
$\us_{\dami}$, conditional on the value of $\s_i$. The correct
probability law for this generation reads
\begin{equation}
\mu(\us_{\dami}|\s_i) = \frac{1}{z} \exp\left[-\beta
E_a(\s_i,\us_{\dami}) + \beta \sum_{j \in \dami} h_{j \to a} \s_j
\right] \ ,
\end{equation}
where $z$ stands for a normalization constant. Once all the
variables at distance 1 from $i$ have been fixed, one can proceed in
the same way with variables at distance 2, and so on and so forth
until the complete configuration $\us$ has been generated.

We can now wrap together the recursive computation of the
conditional magnetizations and the recursive broadcast generation of
the reference equilibrium configuration. We shall indeed define
$P_{i \to a}^{\s_i}(h'_{i \to a})$ as the (density of) probability
that $h_{i \to a}^{\us_B}$ is equal to $h'_{i \to a}$ when $\us_B$
is part of an equilibrium configuration $\us$ drawn conditional on
the value of $\s_i$. A similar definition holds for $Q_{a \to
i}^{\s_i}(u'_{a \to i})$; one can derive recursion relations between
these probability distributions,
\begin{eqnarray}
P_{i \to a}^{\s_i}(h'_{i \to a}) &=& \int \prod_{b \in \dima} \dd
Q_{b \to i}^{\s_i}(u'_{b \to i}) \ \delta\left(h'_{i \to a}- \sum_{b
\in \dima} u'_{b \to i}\right) \ ,
\label{eq_P} \\
Q_{a \to i}^{\s_i}(u'_{a \to i}) &=& \sum_{\us_{\dami}}
\mu(\us_{\dami}|\s_i) \int \prod_{j \in \dami} \dd P_{j\to
a}^{\s_j}(h'_{j \to a}) \\ & & \hspace{3cm} \delta\left(u'_{a \to i}
-f_{a \to i}(\{h'_{j \to a}\}_{j \in \dami}) \right) \ , \nonumber
\label{eq_Q}
\end{eqnarray}
complemented with the boundary condition $P_{i \to a}^{\s_i}(h'_{i
\to a}) = \delta(h'_{i \to a} - \s_i \infty)$ when $i \in B$. From
the solution of these recursion equations one finally obtains
\begin{eqnarray}
G(i,B) &=& \sum_{\s_i} \s_i \frac{e^{\beta \s_i H_i}}{2 \cosh(\beta
H_i)} \int \prod_{a \in \di} \dd Q_{a \to i}^{\s_i}(u'_{a \to i})
\tanh\left(\beta \sum_{a \in \di} u'_{a \to i}\right) \nonumber \\
&&- \tanh^2(\beta H_i) \ . \label{eq_pts_fg}
\end{eqnarray}

The computation of the point-to-set correlation function we just
presented is exact for any model whose factor graph is a finite
tree. This is not precisely the case of problems on random graphs,
as for instance the diluted $p$-spin model defined by
Eqs.~(\ref{eq_def_pspin},\ref{eq_def_Jdiluted}). In such models the
factor graph is only locally tree-like, loops do exist but most of
them have a length which diverge in the thermodynamic limit. In
consequence the properties of random graph models can be inferred
from such computations on finite trees, provided one takes into
account the presence of the loops as self-consistent boundary
conditions. This kind of reasoning has been extensively developed in
the recent years under the name of the cavity
method~\shortcite{MePa_Bethe,MeMo_book}, which is equivalent to the
replica method~\shortcite{Mo_rep_dil} yet more practical for diluted
systems. Following this methodology one can perform the average of
the point-to-set correlation function (\ref{eq_pts_fg}) with respect
to the quenched disorder of the model. Without entering the details,
this leads to functional recursion equations of the form
$P_{\ell+1}=F[P_\ell]$, where $P$ is a probability distribution over
real numbers (the $h'$ in (\ref{eq_P})), and the recursion is over
the radius $\ell$ of the ball on which the correlation is computed.
Numerical algorithms to deal with such equations between probability
distributions have been extensively used to solve the equations of
the cavity method. The idea, known in this context as population
dynamics~\shortcite{MePa_Bethe}, is to approximate a probability
distribution by a sample of elements, with an empirical distribution
as close as possible to the true distribution.

\begin{figure}
\hspace{-1.5cm}
\includegraphics[width=6.8cm]{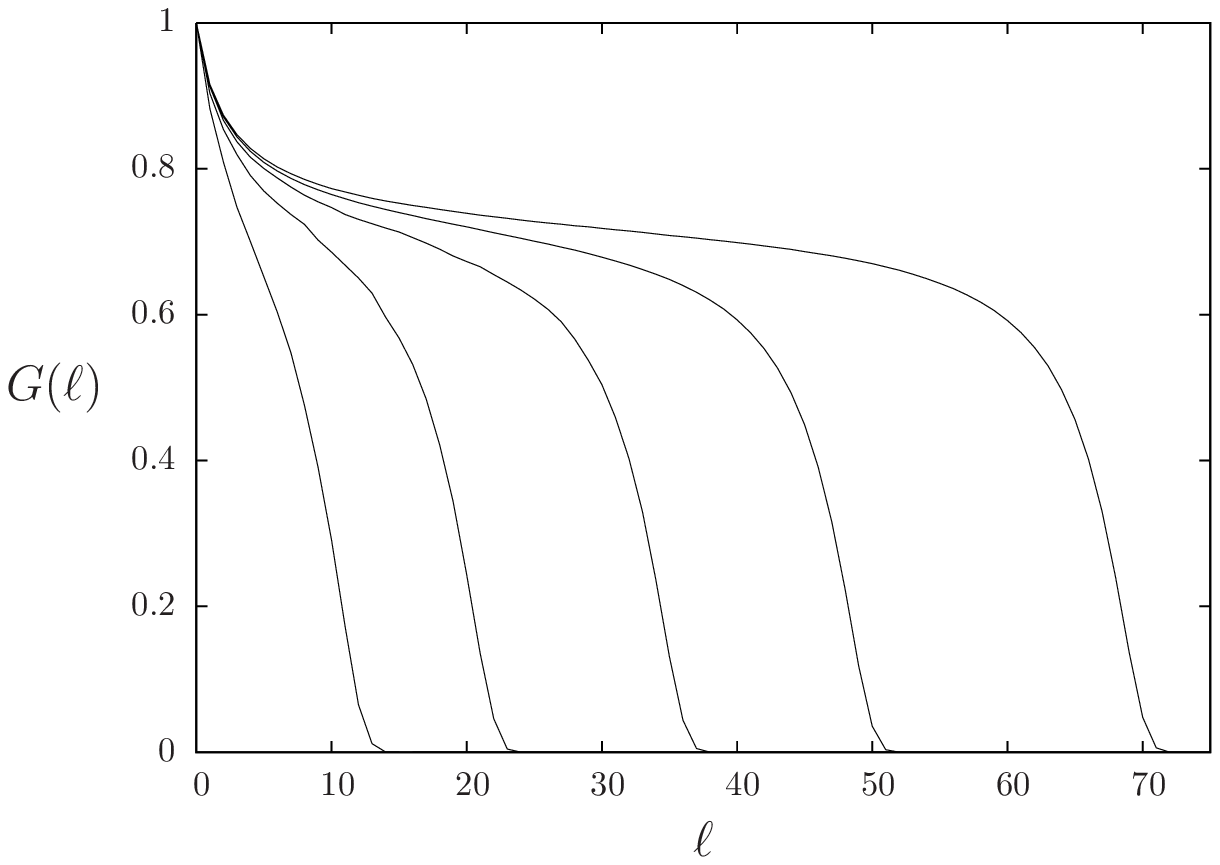} %\hspace{0.5cm}
\includegraphics[width=7.cm]{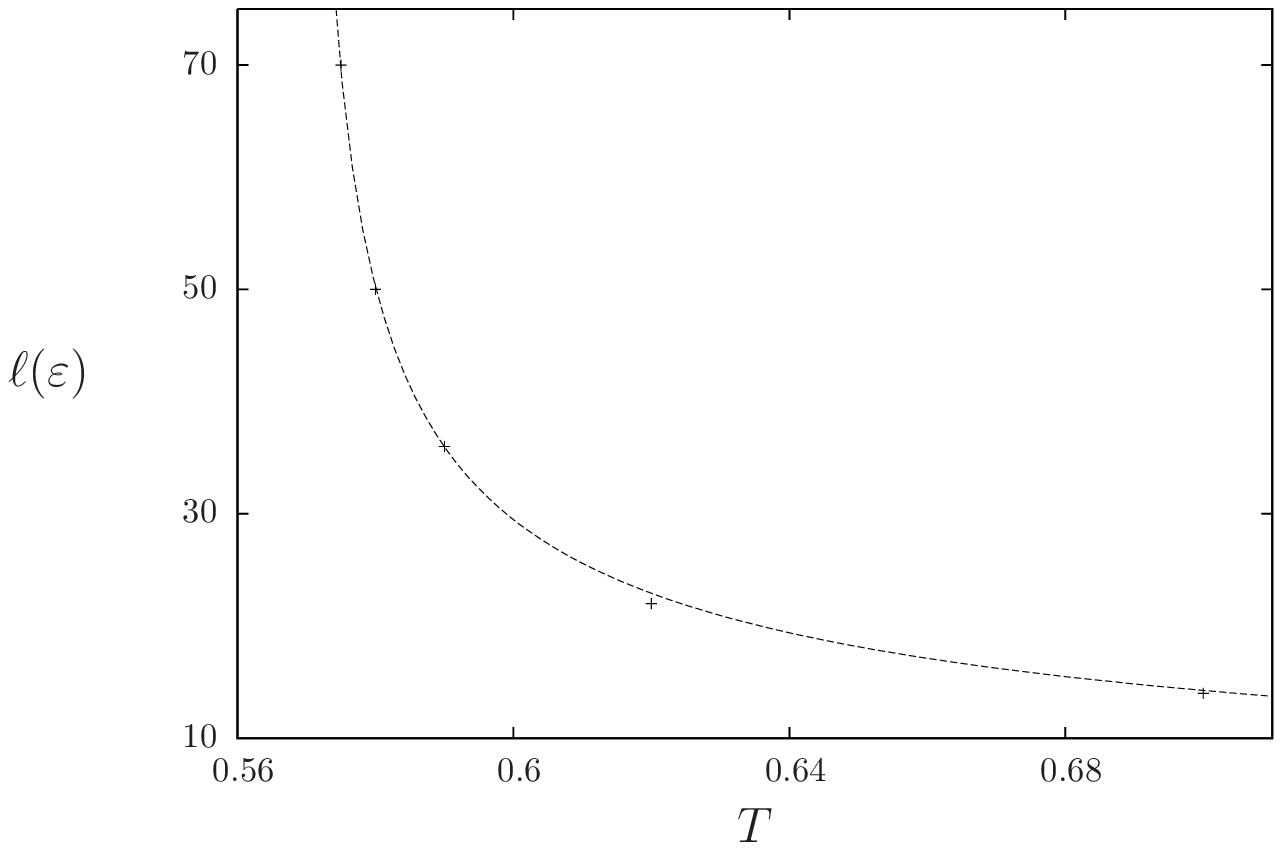}
\caption{Left: Point-to-set correlation function for the $p=3$
diluted mean-field model with $\alpha=1$, from left to right,
$T=0.7$, $0.62$, $0.59$, $0.58$ and $0.575$. Right: the point-to-set
correlation length with $\ve=0.1$, the dashed line is a fit to this
data according to the asymptotic expression $\ell_*(\ve) = C\,
(T-T_{\rm d}(\alpha))^{-1/2}$, with $T_{\rm d}(\alpha=1)=0.51695$. }
\label{fig_plot_pts_diluted}
\end{figure}

\begin{figure}
\centerline{\includegraphics[width=8.cm]{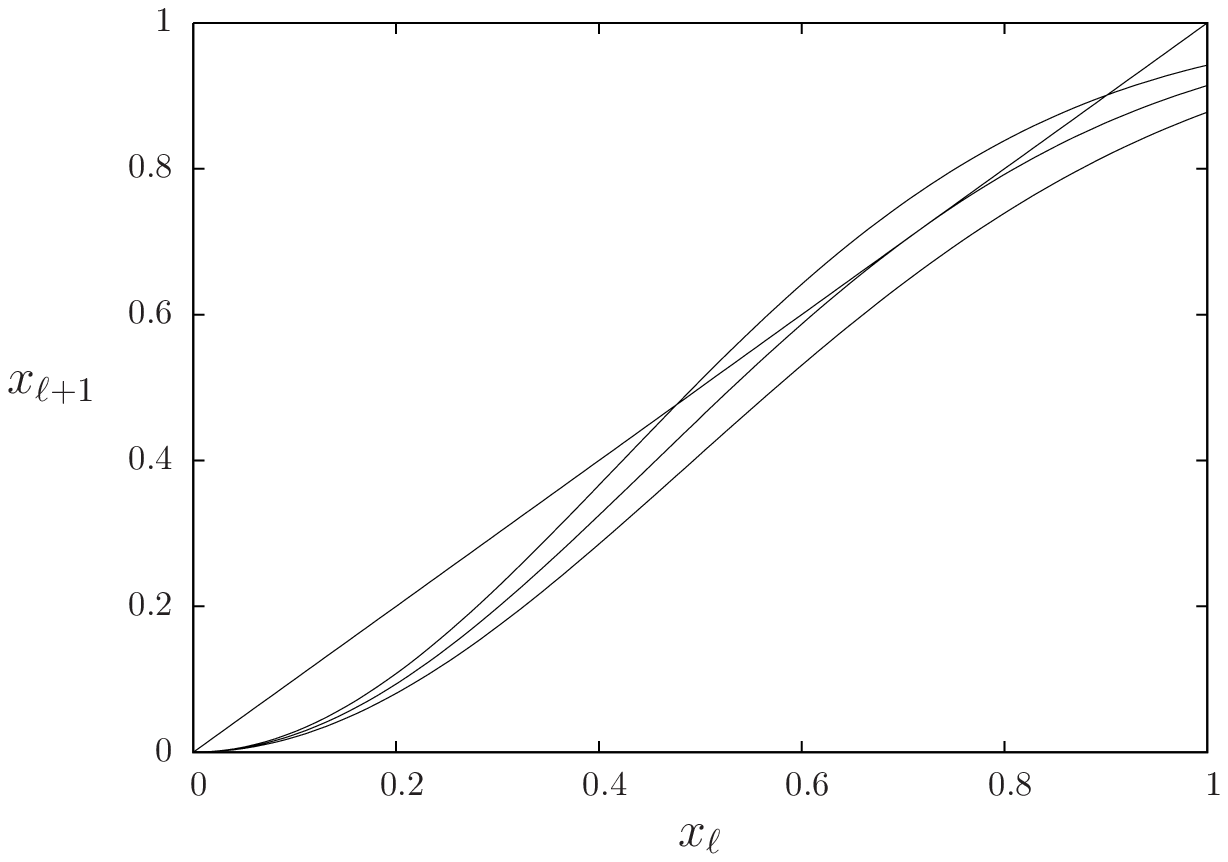}} \caption{A
representation of the bifurcation for the equation $x_{\ell
+1}=f(x_\ell)$. From bottom to top the parameters in $f$ are in the
liquid phase, precisely at the transition, in the glass phase.}
\label{fig_toy_div}
\end{figure}

We show in the left panel of Fig.~\ref{fig_plot_pts_diluted} the
results of such a computation for the average point-to-set
correlation, as a function of the radius of the ball, in the case of
the $p=3$ diluted model. The density parameter $\alpha=1$ is larger
than the zero-temperature limit $\alpha_{\rm d}\approx 0.818$ of the
dynamic transition line (recall the sketch of
Fig.~\ref{fig_plot_sketch_dilpspin}). Indeed one clearly sees a
plateau developing a diverging length upon decreasing the
temperature towards $T_{\rm d}(\alpha)>0$. As discussed above the
presence of this plateau makes the correlation length $\ell(\ve)$
asymptotically independent of the precise choice of $\ve$, as long
as it is smaller than the plateau value. This length is plotted in
the right panel of Fig.~\ref{fig_plot_pts_diluted} as a function of
temperature. As the dashed line shows the divergence of the length
occurs with a critical exponent $1/2$, i.e. $\ell(\ve) \sim
(T-T_{\rm d}(\alpha))^{-1/2}$. One can expect this exponent to be
generic for almost all random graph models with such a discontinuous
transition. It is indeed a natural consequence of the recursive
structure of the equation underlying the computation, namely
$P_{\ell+1}=F[P_\ell]$. In the liquid phase the only fixed point of
the equation $P=F[P]$, reached when $\ell \to \infty$, corresponds
to a vanishing correlation. When the parameters $(\alpha,T)$ are
taken to the glassy phase a bifurcation in this fixed point equation
occurs, a non-trivial solution with a strictly positive correlation
appears discontinuously. To simplify the discussion let us consider
a toy model where the recursion is upon a real $x$ instead of a
function $P$, $x_{\ell +1}=f(x_\ell)$ (this corresponds actually to
the zero temperature limit of the diluted $p$-spin model,
see~\shortcite{MoSe2} for more details). The behaviour of the
function $f$ upon approaching such a discontinuous bifurcation is
shown in Fig.~\ref{fig_toy_div}. It is then a simple exercise to
expand the function $f$ when its parameter is slightly in the liquid
phase, and its argument $x$ is around the value of the close-by
bifurcation, and to show in such a way that the length of the
plateau in $x_\ell$ does indeed diverge with an exponent $1/2$. The
functional nature of the recursion in the general computation should
not spoil this argument, one can thus expect to find generically
such an exponent for the divergence of the point-to-set correlation
length in mean-field diluted models.

An important point should be emphasized here~: for any kind of
transition studied on a Bethe lattice where a correlation length
(shortest path on the tree) diverges as $\epsilon^{-\nu_{\rm tree}}$
($\epsilon \to 0$ controlling the approach to the critical point),
the mean-field exponent of the transition should be deduced as
$\nu_{\rm mf} = \frac{\nu_{\rm tree}}{2}$ and not $\nu_{\rm tree}$.
Indeed the finite connectivity Bethe lattice should be pictured as
embedded in an hypercubic lattice of very large dimension $d$. In
this limit a path of length $\ell$ along the tree corresponds to a
random walk (almost independently of the other branches of the tree)
on the embedding lattice, with an end-to-end separation of order
$\sqrt{\ell}$ in Euclidean distance~\footnote{The validity of this
argument, which is found for instance in~\shortcite{argnu1,argnu2},
can be checked on the simple example of a particle freely diffusing
on a Bethe lattice.}. The mean-field value of $\nu$ for the
transition discussed here is hence $1/4$, we shall come back on this
point in Sec.~\ref{mathKac}.

\subsection{Numerical simulations of diluted mean-field models}

The mean-field diluted models provide an ideal playground to assess
the relevance of the bounds (\ref{eq_bound}) between correlation
lengths and correlation times: we have seen above how to compute the
correlation length in the liquid phase and its divergence at the
dynamical transition temperature $T_{\rm d}$. In addition the
dynamics of these models can be easily investigated with Monte Carlo
simulations (for out-of-equilibrium dynamics see for
instance~\shortcite{BaZe,MoRi}). The equilibrium correlation times
have been thus determined in~\shortcite{MoSe}, for a variation of
the diluted $p$-spin model with constant degrees for the spins, and
allowed to justify the claim of almost optimality of
(\ref{eq_bound}) without additional hypotheses. Indeed for $T>T_{\rm
d}$ one finds that the equilibrium correlation time remains finite
in the thermodynamic limit, and diverges algebraically when $T \to
T_{\rm d}^+$. Hence the lower bound of (\ref{eq_bound}) could only
been improved to include a dynamical exponent $z$ and be promoted to
$\tau \ge \ell^z$, with probably $z\ge 2$ for a large class of
models. On the contrary in the low temperature phase, $T<T_{\rm d}$,
it is the upper bound of (\ref{eq_bound}) which is saturated (up to
an improvement of the constant $C_2$): the equilibrium correlation
time diverges exponentially with the size $N$ of the system, and the
correlation volume is itself proportional to $N$.

Coming back to the diluted $p$-spin model with Poissonian degrees,
let us mention a particular feature which is apparent in the sketch
of Fig.~\ref{fig_plot_sketch_dilpspin}. The dynamic transition line
$T_{\rm d}(\alpha)$ reaches the zero temperature axis at a finite
value $\alpha_{\rm d}$. The neighborhood of the point
$(\alpha=\alpha_{\rm d},T=0)$ thus exhibits a rich crossover
phenomenon between low temperature activated dynamics controlled by
energy barriers and positive temperature transitions of the
schematic Mode Coupling Theory kind (which is an exact description
of the $\alpha \to \infty$ limit~\shortcite{MCT_pspin}). We refer
the reader to~\shortcite{MoSe2} for more details.

\subsection{Connection with the replica symmetry breaking formalism}
\label{connection}

We shall close this section on mean-field diluted systems with a
discussion of the relationship between the computation of the
point-to-set correlation function and the usual cavity formalism at
the level of one step of replica symmetry breaking (1RSB), first
unveiled in~\shortcite{MeMo}. The existence of such a connection
should be expected by the reader accustomed to the fully-connected
$p$-spin model: we argued that in the phase diagram sketched in
Fig.~\ref{fig_plot_sketch_dilpspin} the transition line $T_{\rm
d}(\alpha)$, defined as the point where the point-to-set correlation
length diverges, can be meaningfully called the dynamic transition.
Indeed the lower bound in Eq.~(\ref{eq_bound}) implies a divergence
of the equilibrium correlation time on this line. However the large
$\alpha$ limit of the diluted $p$-spin model gives back the
fully-connected one. For the latter the dynamic transition has been
studied since a long time, and its appearance is known to be related
to the static properties of the model studied by various methods, in
particular the replica and the TAP ones. One of these
characterizations is the existence of a non-trivial solution of the
1RSB equations with the Parisi breaking parameter $m=1$.

One can indeed recover this equivalent definition of the dynamic
transition in generic mean-field diluted systems. Let us come back
on the computation on a finite tree presented in
Sec.~\ref{sec_dil_computation}. We introduced there the distribution
$P_{i \to a}^{\s_i}(h'_{i \to a})$ as the probability of $h_{i \to
a}^{\us_B}$ with respect to the choice of the boundary configuration
$\us_B$ conditioned on $\s_i$. One can instead define the
unconditional version of this distribution, $P_{i \to a}(h'_{i \to
a})$, which is related to the conditional one by
\begin{equation}
P_{i \to a}(h'_{i \to a}) = \sum_{\s_i} \frac{e^{\beta h_{i \to a}
\s_i}}{2 \cosh(\beta h_{i \to a})} P_{i \to a}^{\s_i}(h'_{i \to a})
\ .
\end{equation}
A similar definition holds for $Q_{a \to i}(u'_{a \to i})$. A
computation whose details can be found in~\shortcite{MeMo} shows
that the conditional equation (\ref{eq_P}) translates into
\begin{eqnarray}
 P_{i \to a}(h'_{i\to a})& =& \frac{1}{{\mathcal Z}_{i \to a}} \int \prod_{b \in \dima}
\dd Q_{b \to i}(u'_{b \to i}) \
\delta\left(h'_{i \to a}- \sum_{b \in \dima} u'_{b \to i}\right) \ \nonumber\\
&& \times\left( \frac{\cosh(\beta h'_{i \to a})}{\underset{b \in
\dima}{\prod} \cosh(\beta u'_{b \to i})} \right) \ .
\end{eqnarray}
This is nothing but the usual 1RSB equation with
$m=1$~\shortcite{MePa_Bethe} (in general the last term in
parenthesis, called reweighting factor or free-energy shift, is
raised to the power $m$). A similar observation holds for the
equation on $Q_{a \to i}$ obtained from (\ref{eq_Q}). The 1RSB
equations are usually solved on a random factor graph without any
explicit boundary $B$. In consequence the existence of a non-trivial
solution of the usual 1RSB equations with $m=1$ is equivalent to the
presence of long-range point-to-set correlations (with a fictitious
boundary sent to infinity).

This computation provides also an alternative interpretation of the
1RSB equations for mean-field diluted systems. $P_{i \to a}$ is
usually considered as the distribution of an effective field with
respect to the choice of a \emph{pure state} of the system. This can
be intuitively defined for a random graph of large but finite size
as a portion of the configuration space sufficiently disconnected
from the other pure states, yet a clear-cut mathematical definition
is not easy to handle. One can alternatively interpret $P_{i \to a}$
as a distribution over the boundary conditions of a tree of very
large depth. This is precisely the computation done for the
point-to-set correlation, i.e. the 1RSB equations with $m=1$, where
the boundary condition is drawn itself from the Gibbs-Boltzmann
distribution. One can moreover generalize this interpretation to
arbitrary values of $m$, biasing accordingly the probability
distribution of the boundary conditions~\shortcite{MoRiSe}.

\section{Kac models}
\label{Kacmodels}

% In the previous section we have seen how spin models on diluted random
% graphs can be used to study correlation lengths and the effect of
% finite interaction range.  This has allowed to exhibit an example on
% how the point-to-set (PS) correlation length grows and diverge
% together with the relaxation time as the temperature decreases towards
% the dynamical transition.  Unfortunately, these models, which are
% solved exactly within Mean Field Theory, cannot help in resolving
% issues related to mean field pathologies. As for long range models
% they present infinite life metastable states, and non-convex
% free-energy.

% In this section we would like to discuss finite dimensional models
% with disordered Kac interactions.

A different class of systems in which the computation of PS
correlation is possible is provided by Kac models. Kac models are
classical tools of statistical physics to study the relationship
between finite dimensional systems and their Mean Field
counterparts~\shortcite{Kac,LP1,LP2}.  In these models one considers
variables interacting through a potential of growing range $r_0$ and
intensity decreasing in such a way as to keep finite the total
interaction strength of one particle with the surrounding
environment. For any finite values of $r_0$ the general properties
of finite dimensional system such as convexity of free-energy and
absence of infinite life metastable states hold.  There properties
break down in mean-field thoery, which holds exactly in the limit of
weak long range forces in which $r_0$ scales as the linear system
size $L$ and both lengths tend to infinity together. The relation
between the two regimes can be understood studying the so-called Kac
limit~\shortcite{Kac}, where one considers a large interaction range
$r_0\to \infty$ but only after having taken the thermodynamic limit
$L\to\infty$.  Remarkably one finds that while the mean-field
predictions hold in single phase regions, some of the typical
pathologies of mean-field theory, such as the non convex
free-energies are removed and the Maxwell construction emerges
naturally. The resulting theory is an improved mean field theory
that incorporarate the spatial dimension, with the possibility of
describing inhomogeneous configurations and interfaces.  For some
systems (see, for instance, Refs.~\shortcite{Bov,Leb}) it has been
possible to devise asymptotic expansions around $r_0=\infty$ which
have led to the analysis of decay of metastable states or proofs of
the existence of phase transitions. It is natural in the context of
glassy systems to address the study of correlation lengths in the
Kac limit. This provides a solvable limit case to confront with
phenomenological theories based on mean-field
ideas\shortcite{KTW,BiBo} and gives rise to a detailed picture with
two relevant length scales.

\subsection{Definition}
 \label{mathKac}

Disordered models with Kac interactions were first described
in~\shortcite{FZ}, the case of spin glass models with pair
interactions. Following the general philosophy of this chapter, here
we are interested in more general models with $p$-body interactions
providing a minimal finite dimension modification of the fully
connected $p$-spin model (\ref{eq_def_pspin}).

We consider then spins $\sigma_i$ in the $d$ dimensional hypercube
$i\in \Lambda=\{1,...,L\}^d$ with periodic boundary condition. The
variables $\sigma_i$ can be of various nature, for example they can
be Ising variables $\sigma_i=\pm 1$ or real variables subject to
some local spherical constraint as we will specify below. The Kac
$p$-spin Hamiltonian is defined by~\shortcite{FTjpa}
\begin{eqnarray}
E({\us})=-\sum_{i_1, i_2,...i_p\in \Lambda}
J_{i_1,...,i_p}\sigma_{i_1}...\sigma_{i_p} \label{pspin}
\end{eqnarray}
where, as for the mean-field model, the coupling constants $
J_{i_1,...,i_p}$ are independent centered Gaussian variables,  but
now differently from the mean field model, the variance depends on
the mutual distances between the indexes according to:
\begin{eqnarray}
\mathbb{E}[ J_{i_1,\dots,i_p}^2]=\frac {1}{r_0^{pd}} \sum_{k\in
\Lambda} \psi\left(\frac{|k-i_1|}{r_0}\right)\dots
\psi\left(\frac{|k-i_p|}{r_0}\right) \ . \label{psipsi}
\end{eqnarray}
The function $\psi(x)$ can be chosen somehow arbitrarily, provided
it fulfills the following constraints:
\begin{itemize}
\item it is positive, $\psi(x)\geq 0$.
\item it becomes negligible when its argument is much larger than 1.
\item it is normalized such that $\int d^dx \;\psi(x)=1$.
\end{itemize}
These conditions guarantee that only spins within distances of order
$r_0$ have an effective direct interaction. The function $\psi$ can
also be used to define a locally spherical model that without
changing the physics from the Ising case, allows important
simplifications in analytic studies. This is done considering
$\sigma_i\in \mathbb{R}$ subject to the conditions
$\frac{1}{r_0^d}\sum_{j\in\Lambda} \psi\left(\frac{|i-j|}{r_0}
\right)\s_i^2=1$ in all points of space $i$.
% Possible choices of $\psi(x)$ are $\psi(x)=C
% \I_{|x|<1}$ or $\psi(x)=C' \exp\left(-|x|\right)$ with $C$ and $C'$
% suitable constants that enforce normalization.
Note that the disorder distribution implies that the Hamiltonian
itself has a Gaussian distribution with covariance specified by
\begin{eqnarray}
\mathbb{E}[E(\us^{(1)}) E(\us^{(2)})]= \sum_i
Q_i(\us^{(1)},\us^{(2)})^p \ , \label{pspin2}
\end{eqnarray}
where $Q_i(\us^{(1)},\us^{(2)})=\frac{1}{r_0^d} \sum_{j} \psi\left(
\frac{|i-j|}{r_0} \right) \sigma_j^{(1)}\sigma_j^{(2)}$ measures the
similarity of the two configurations $\us^{(1)}$ and $\us^{(2)}$ on
a ball of scale $r_0$ around $i$.  Models that compound different
values of $p$ can be defined considering gaussian Hamiltonian with
correlations $ \mathbb{E}[E(\us^{(1)}) E(\us^{(2)})]= \sum_i
\phi(Q_i(\us^{(1)},\us^{(2)})) \ . $ The function $\phi(q)$ should
satisfy suitable conditions to define a non-negative covariance
matrix.

Following the original suggestion by Kac~\shortcite{Kac}, the
scaling of the interaction with $r_0$ is chosen in such a way to
insure the existence of the limit of the free-energy for
$r_0\to\infty$. Different ways of taking this limit allows to study
the relation between systems with finite range interactions and
their mean field counterpart.  If $r_0$ is chosen to scale with the
system size $r_0=L\gg 1$ we have back the mean-field limit. All
groups of $p$ spins interact and the lattice structure and the space
dimension become completely inessential. But one can consider a
different procedure, called Kac limit, where $r_0\to\infty$ only
after the thermodynamic limit $L\to\infty$. A requirement for Mean
Field to have a qualitative relevance in the description of finite
range systems is a smooth $r_0$ cross-over from $r_0\ll L$ to
$r_0\sim L$ ($L\gg1$).  Rigorous results connecting the Kac limit to
the mean-field limit have been proved in \shortcite{FTprl,FTjpa},
under the technical hypothesis of parity of the function $\phi$,
$\phi(q)=\phi(-q)$. These findings can be informally and
qualitatively stated as follows:
\begin{itemize}
\item The free-energy of the Kac model tends for all temperature to
  the free-energy of the corresponding mean-field model in the Kac
  limit.
\item In the same limit, the structure of the correlation functions on
  length scales of order of $r_0$ approaches the structure of
  mean-field correlations.
\end{itemize}
The first point is a continuity property, it tells that the mean
field limit is a good starting point in the evaluation of the
free-energy for large $r_0$. Of course this does nor necessarily
imply that the phase transitions found in mean field are also
present for some finite $r_0$. For example, it is clear that in 1D
the free-energy is an analytic function of temperature for all
finite $r_0$, despite the fact that the large $r_0$ limiting
function exhibit a phase transition.

The second point clarifies that mean-field approximation is at least
capable to capture the local properties of the system. This is
particularly interesting at low temperature where it implies that
the system behaves as an ideal glass at least locally on length
scales of order $r_0$. The possibility that this description extends
to scales much
 larger than $r_0$ in some finite space dimension remains an open question.
To go beyond these general rigorous results and study quantitatively
the behavior of correlation lengths, tools of theoretical physics
and non-rigorous techniques are needed as it will be explained in
the next sections.

\subsection{Effective potential: a Landau free-energy functional for glasses}
\label{landau}

In this section we would like to review the basic tool of the {\it
glassy effective potential}~\shortcite{FPpot}, which has been
instrumental in the computation of  point-to-set correlation lengths
in the Kac model \shortcite{FranzMontanari}.

The effective potential is a Landau kind of free-energy as function
of an order parameter explicitly devised to be sensitive to stable
or metastable glassy states.  While this construction is in
principle very general, for sake of simplicity we first expose it in
the case of fully connected models, and later explain how to extend
the analysis to Kac models.  The basic idea is that metastable
states can be studied through restrictions of the Boltzmann-Gibbs
measure to the surroundings of a representative configuration. One
then observes that in glassy systems where equilibrium is composed
by an extensive number of metastable states, any equilibrium
configuration can be taken as a representative configuration. Let us
then choose a random configuration $\us^0$ according to the
equilibrium measure, and consider a second system $\us$ which is in
a constrained equilibrium where the overlap with the reference
configuration $q(\us^{(0)},\us)=\frac 1 N \sum_i \s_i\s_i^0$ is
fixed to some preassigned values $p$.\footnote{Note that we used
already the letter $p$ to define the $p$-spin Hamiltonian. It will
be clear from the context to which quantity the notation refers.}

The free-energy of this second system, for fixed $\us^{(0)}$ is
\begin{eqnarray}
&&F[p,{\us^{(0)}}]=-\frac{T}{N} \log Z[p,\us^{(0)}]\\
& &Z[p,\us^{(0)}]=\sum_{\us}  e^{-\beta E(\us)}\prod
\delta(q(\us^{(0)},\us)-p).
\end{eqnarray}
In the thermodynamic limit this is
% The conditional probability of $\s$ given $\s^{(0)}$ is
% \begin{eqnarray}
% & &\mu[\us|\us^{(0)},p]=\frac{1}{Z[p,\us^{(0)}]}  e^{-\beta E(\us)}
% \delta(q(\us^{(0)},\us)-p).
% \end{eqnarray}
% where
% \begin{eqnarray}
% & &Z[p,\us^{(0)}]=\sum_{\us}  e^{-\beta E(\us)}\prod
% \delta(q(\us^{(0)},\us)-p).
% \end{eqnarray}
% The corresponding constrained free-energy
% $F[p,{\us^{(0)}}]=-\frac{T}{N} \log Z[p,\us^{(0)}]$
is self-averaging with respect to the quenched disorder in the
interactions and the choice of the equilibrium configuration
$\us^0$,
\begin{eqnarray}
& &F[p]=\mathbb{E} \langle F[p,\us^{(0)}]\rangle \ , \label{effpot}
\end{eqnarray}
where by $\langle\cdot\rangle$ we denoted the average with respect
to the equilibrium distribution for $\us^{(0)}$ and by $\mathbb{E}$
the average with respect to the quenched disorder.  The function
$F[p]$ can be used as a Landau free-energy as a function of the
order parameter to detect and study glassy states. It is customary
to define the effective potential $V[p]$ subtracting the
unconstrained value of the free-energy, $V[p]=F[p]-F$. This is the
large deviation function allowing to evaluate the probability
$\P[p]$ of the overlap with a reference configuration
\begin{eqnarray}
\P[p]=\exp\left( -\beta V[p]\right) \ .
\end{eqnarray}
The analysis of the effective potential, which requires the
computation of the double average in (\ref{effpot}) can be performed
in mean-field theory through the replica method, or alternatively
the cavity method. For the family of models we are interested in,
the most general ansatz needed to describe $V[p]$ is of a 1RSB form,
that accounts for the possibility of having multiple metastable
states.  In fully connected models this takes a simpler form with
respect to the diluted models described in section \ref{connection}.
This is just parametrized by the Parisi breaking parameter $m$ and
two real variables $q_1$ and $q_0$, which represent respectively the
typical value of the overlap between configuration in the same
metastable state and in different ones \shortcite{MPV}.  The
parameter $m$ can be tuned to select different families  of
metastable states with different free-energies. If the equilibrium
phase is composed by an extensive number of metastable state,
corresponding to a non-zero configurational entropy $m$ takes the
value $m=1$.

\begin{figure}[ht]
\begin{center}
\includegraphics[width= 0.7 \textwidth]{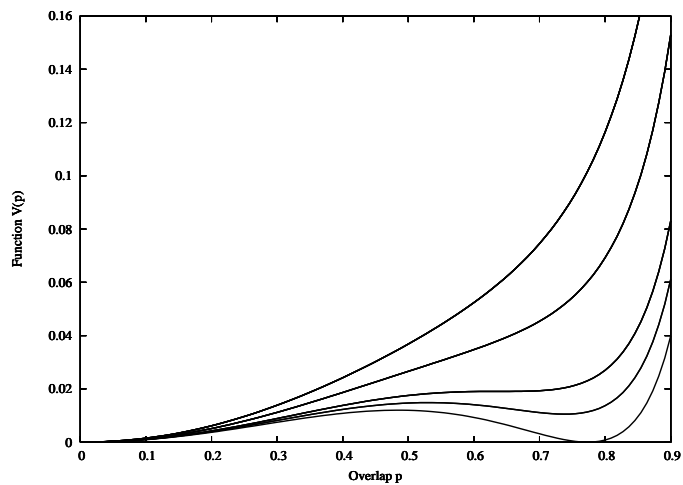}
\caption{The function $V(p)$ at different temperatures. For
comparison with the case of Kac model we consider a Hamiltonian with
two body and 4 body interactions with $\phi(p)=1/2(0.1 \times
p^2+p^4)$.  From top to
  bottom $T= 0.703486 > T^*$, $T=T^*=0.633137$, $T=T_{\rm d}=0.57525$,
  $T=0.558049$, $T=T_{\rm c}=0.541847$. The function is convex for
  $T>T^*$. It has an inflection point with positive slope for
  $T_{\rm d}<T<T^*$. In the interval $T_{\rm c}<T<T_{\rm d}$,
  $V(q)$ has a local minimum
  for a temperature dependent value $p=q_{EA}$. The difference
  $V(q_{EA})-V(0)$ is ($T$ times) the bulk configurational entropy
  $V(q_{EA})-V(0)=T\Sigma_\infty(T)$.  }
\label{fig:pot}
\end{center}
\end{figure}

In the spherical p-spin model, defined by the Hamiltonian
(\ref{eq_def_pspin}), the effective potential can be computed in a
closed form as a saddle point over $q_1$, $q_0$:
\begin{eqnarray}
&&V[p]={\max_ {q_1,q_0}}\, {\mathcal V}(p,q_1,q_0,m) \\
&&{\mathcal V}(p,q_1,q_0,m) =
%\left(
-\frac 1 2 \beta [2 \phi( p) - m \phi(q_0) - (1 - m) \phi( q_1)] +
% \right.
\nonumber \\
 &&
\frac {1}{2\beta} \left( \frac{p^2 - q_0} {1 - (1 - m) q_1 - m q_0
}+
% \right.
\nonumber
%  &&
% \left.
 \frac{1 - m}{m}\log[1 - q_1]-  \frac 1 m \log[1 - (1 - m) q_1- m q_0] \right)
\label{potmf}
\end{eqnarray}
where we denoted $\phi(q)=\frac 1 2 q^p$.  The shape of the function
$V[p]$, which depends on temperature, is sensitive to the existence
of metastable states.  Indeed, as can be seen in Fig.~\ref{fig:pot},
one can identify three model dependent temperatures of interest
where qualitative changes take place.  Two of them are the
mean-field dynamical transition temperature $T_{\rm d}$ and the
static transition temperature $T_{\rm c}$. In addition there is a
third temperature $T^*$, with $T^*>T_{\rm d}>T_{\rm c}$, first
identified in~\shortcite{FPpot}.  Above $T^*$ the function $V$ is a
convex function with a single minimum at $p=0$. At $T^*$ and
inflection point appears, and below that temperature the potential
is non-convex. For temperature between $T_{\rm d}$ and $T^*$, the
function continues to have a single minimum for $p=0$.  At $T_{\rm
d}$ a local minimum at a value $p=q_{EA}>0$ develops. In the
interval $[T_{\rm c},T_{\rm d}]$, the point $p=0$ is still the
absolute minimum of $V$.  The two minima structure below $T_d$
reflects the partition of the equilibrium measure in disjoint
metastable states.  The value $q_{EA}$ is the typical overlap
between configurations belonging to the same metastable state. For
$p=q_{EA}$ $\us$ is in the state specified by $\us^{(0)}$. Different
metastable states have zero mutual overlap. For $p=0$ all but the
metastable state specified by $\us^{(0)}$ contribute to the
free-energy and $V(0)=0$. Correspondingly, the difference in
free-energy between the two minima equals the system's
configurational entropy $\Sigma_\infty(T)$ multiplied by
temperature. The configurational entropy vanishes linearly on
approaching $T_{\rm c}$,
 $\Sigma_\infty(T)\sim T-T_{\rm c}$ and the two minima
become degenerate. Below that temperature the mean field model is in
an ideal glassy state and the two minima remain degenerate.

We now generalize the previous construction to Kac models.  As
before,
 we use the overlap with
an equilibrium reference configuration $\us^{(0)}$ as an order
parameter.  In this case however, we are interested in considering
the free-energy as a functional of an order parameter profile that
is space dependent. In view of considering large values of the
interaction range $r_0$ it is natural to coarse-grain the order
parameter on scales $\delta$ such that $\delta\simeq dx \; r_0 \gg
1$ with $dx \ll1$~\shortcite{presutti,FTJSTAT}. This allows to
rescale all lengths $x=i/r_o$ which are then measured in units of
the interaction range $r_0$. Let us denote $q_x(\us^{(0)},\us)=\frac
{1}{\delta^d}\sum_{i\in B_{x} }\s^{(0)}_i\s_i $ the local overlap in
a square cell $B_x$ of linear size $\delta$ labeled by $x$ between
two configurations $\us^{(0)}$ and $\us$.  In complete analogy with
the long-range case, one can define an effective potential
functional $W[p(x)]=F[p(x)]-F$ as the average free-energy excess for
imposing an overlap profile $p(x)$.
\begin{eqnarray}
& &F[p(x)]=-\frac{T}{N}\mathbb{E}  \log Z[p(x),\us^{(0)}]\\
& &Z[p(x),\us^{(0)}]=\sum_{\us}  e^{-\beta E(\us)}\prod_x
\delta(q_x(\us^{(0)},\us)-p(x)). \label{uva}
\end{eqnarray}
We can now relate the point-to-set correlation function to the
effective potential. Consider the  correlation function
 between point $0$ and the set $B$
considering profiles such that $p(x)=1$ is fixed for $x\in B$.
Observing that the probability of a profile can be written as
$\mathbb{P}[p(x)]=\exp(- W(p(x)))$ we find
\begin{eqnarray}
\langle \langle p(0) \rangle_{\us^{(0)}_B} \rangle =
\frac{\int_{\{p(x)=1\; ;\; x\in B \}} {\mathcal D}{p}(x) \;
e^{-\beta W[{p}(x)]} {p}(0)} {\int_{\{ {p}(x)=1\; ;\; x\in B \}}
{\mathcal D}{p}(x) \; e^{-\beta W[{p}(x)]} } \ . \label{p0}
\end{eqnarray}
For general values of $r_0$ the evaluation of the large deviation
functional $W[p(x)]$ is a formidable task. Fortunately, in the Kac
limit $r_0\to\infty$ the computation becomes simpler, it can be
stated as a variational problem that generalize to spatially varying
functions the saddle point evaluation of (\ref{potmf}).  The region
of temperatures $T>T_c$ can be analyzed through the 1RSB theory that
we have discussed in several contexts in this chapter. This
generalizes the description of the fully-connected case in the fact
that the overlap variational parameters become position dependent
functions $q_1(x)$ and $q_0(x)$ (the parameter $m$ remains space
independent). We give the explicit expression of the effective
potential for the p-spin model (\ref{pspin}) with local spherical
constraint, where the dependence on the various parameters can be
written in a closed form
\begin{eqnarray}
&&W[p(x)]={\max_{q_1(x),q_0(x)}} r_0^d \int d^dx\;\left( -\frac 1 2
\beta \{ 2 [\phi( p^\star(x))-\phi(p(x))] - m [\phi( q^\star_0(x))
-\phi(q_0(x))] \right.
\nonumber\\
&& \left.
 - (1 - m) [\phi(
q_1^\star(x)) -\phi(q_1(x))] \} +{\mathcal V}(p(x),q_1(x),q_0(x),m)
\right)  \ , \label{pote}
\end{eqnarray}
% \begin{eqnarray}
% &&W[p(x)]={\max_{q_1(x),q_0(x)}}
% \left\{ r_0^d \int d^dx\;\left(
% -\frac 1 2 \beta [2 (\phi(\psi\star
% p(x))-\phi(p(x))) - m (\phi(\psi\star
% q_0(x))
% -\phi(q_0(x))
% \right.\right.
% \nonumber\\
% &&
%  - (1 - m) (\phi(\psi\star
% q_1(x)) -\phi(q_1(x))) ]
% \nonumber\\
% &&
% \left.
% \left.
% +{\mathcal V}(p(x),q_1(x),q_0(x),m)
% \right) \right\} \ ,
% \label{pote}
% \end{eqnarray}
We have introduced the notation $ f^\star(x)=\int d^dy\; \psi(|x-y|)
f(y)$ for the convolution with $\psi$ and the function ${\mathcal
  V}(p,q_1,q_0,m)$ is given by the mean field expression
(\ref{potmf}). Mutating the terminology from mechanics we will call
action the argument of the max in (\ref{pote}).  We notice that for
profiles that are inhomogeneous on limited spatial extension $W$ is
proportional to the interaction volume $r_0^d$.  The  expression
(\ref{pote}) can be simplified using a (physically harmless) lowest
order gradient expansion for the convolution
$\phi(f^\star(x))=\phi(f(x))+c\phi'(f(x)) \nabla^2 f(x)$
% +\phi''(f(x)) (\nabla f(x))^2$
% ATTENZIONE ALLA GRADIENT EXPANSION
 with $c=\frac{1}{2d}\int
d^dz\; \psi(|z|) z^2$. This approximation does not affect the
physics of the problem while the action takes the familiar
field-theoretical form of the integral of a Lagrangian density,
dependent on the various overlap parameters and their spatial
derivatives.  Notice however that for the expansion to be well
behaved one needs $\phi''(q)>0$ for all $q$. This is not verified
for pure $p$-spin interactions if $p\geq3$ which has $\phi''(0)=0$.
For this reason the numerical computations based on this truncation
have used functions $\phi$ of the kind $\phi(q)=\frac{1}{2}(q^4+a
q^2)$, which correspond to a mixed $p$-spin model with
 two body and four body interactions. For small enough values of $a$
the model has the same phenomenology as the pure $p$-spin. Figures
are presented for the value $a=0.1$.

Thanks to the proportionality of the action to the interaction
volume $r_0^d$, the functional integral in the expression (\ref{p0})
of the PS correlations can be evaluated by saddle points. We will be
then interested in profiles $p(x)$ that are stationary points of the
action $\frac{\delta W }{\delta p(x) }=0$ for $x\in \Bbar$ with
boundary condition $p(x)=1$ for $x\in B$.  These include the
absolute  minima, that represent
 equilibrium
profiles and directly determine the PS correlation function,
 but also relative minima, saddle points and maxima that we will associate
to metastable states and barriers.

The condition of equilibrium (\ref{equal}) discussed in section
\ref{sec_defpts_precise} has interesting implications on the mutual
relation between the overlap functions $q_1(x)$, $q_0(x)$ and
$p(x)$.  Consider the local overlap $q_x(\us',\us'')$, between two
replicas with the same constraint $\us^{(0)}$ on $B$, and its $k$-th
moment $\mathbb{E} \langle q_x(\us',\us'')^k\rangle$. it is possible
to show~\shortcite{MPV}  that within the 1RSB formalism
\begin{eqnarray}
\lim_{r_0\to\infty} \mathbb{E} \langle q_x(\us',\us'')^k\rangle =
(1-m)q_1^k(x)+m q_0^k(x)
\end{eqnarray}
where $q_1(x)$ and $q_0(x)$ are the solution of the maximization
problem (\ref{pote}).  On the other hand, according to
(\ref{equal}),  this expectation value should be simply equal to
$\lim_{r_0\to\infty} \mathbb{E} \langle q_x(\us^{(0)},\us')^k\rangle
=p(x)^k$, thus leading for all $k$ and $x$ to the identity
$(1-m)q_1^k(x)+m q_0^k(x)=p(x)^k$. Two different ways of satisfying
this identities are of physical relevance: 1) a simple solution
describing equilibrium in a single metastable state, such that for
all $x$
\begin{eqnarray}
q_1(x)=q_0(x)=p(x) \ . \label{rs}
\end{eqnarray}
2) A solution with $m\to 1$, for which
\begin{eqnarray}
q_0(x)=p(x) \ ,
\nonumber\\
q_1(x) > p(x) \ . \label{rsb1}
\end{eqnarray}
This solution (see below) is relevant when ergodicity is broken and
the constrained equilibrium is composed by multiple metastable
states with a non-zero configurational entropy $\Sigma$.  According
to general principles~\shortcite{remi} the configurational entropy
can be computed from a solution with the form (\ref{rsb1}) as
$\Sigma=-\beta \frac{1}{m^2} \frac{\partial W}{\partial m}|_{m=1}$.

Non-equilibrium stationary profiles $p(x)$, associated to transition
states and barriers, will be relevant in our analysis.  For these
non-equilibrium solutions, there are no restrictions of principle on
the functions $q_1(x)$ and $q_0(x)$.  Of course the condition
$q_0(x)=q_1(x)=p(x)$ can still be used as an approximation.  Despite
the fact that in some regions this gives rise to clear physical
inconsistencies we believe that it predicts the correct scaling of
the free-energy of metastable and barrier states with temperature
and interaction range.  More complex solutions, studied in
\shortcite{D0} confirm this point.

Let us stick to the forms (\ref{rs}) and (\ref{rsb1}). In both
mentioned cases, one gets great simplifications in the variational
problem one needs to solve. For example for the ergodic equilibrium
profiles (\ref{rs}) the effective potential, using the mentioned
gradient expansion for the convolution reads:
\begin{eqnarray}
&& W[p(x)]= r_0^d\int d^dx\; \left( -\frac{c\beta}{2} ( \phi'(p(x))
\nabla^2 p(x)
%+ \phi''(p(x))(\nabla p(x))^2)
+V(p(x)) \right)
\nonumber\\
&& V(p)=-\frac{\beta}{2} \phi(p) -\frac T 2 \left[ p(x) + \log ({1-
p(x)}) \right]
 \ .
\label{simple}
\end{eqnarray}
In the 1RSB  solution (\ref{rsb1}) an additional $q_1$ dependent
term proportional to $1-m$ appears that allow to compute the
configurational entropy. It is natural to look for solutions
respecting the symmetries imposed by the set $B$.  We will discuss
mainly the spherical geometry, where the set $B$ is just the
exterior of a spherical cavity, $B=\{x: \;  |x|>\ell\}$ and
solutions should only depend on $|x|$. Another interesting choice is
the one of planar geometry, where $B=\{x:\;\;| x_1|>\ell\}$, with
$x_1$ the first component of the $d$ dimensional vector $x$,
solutions should only depend on $x_1$.  In both cases it is
interesting to study the behavior of the solutions as a function of
the cavity size $\ell$, that, remember, we measure in units of $r_0$
as all the lengths. Physical saddle points should be such that in
both cases the profile is flat in the center of the cavity $\nabla
p(0)=0$.  In both geometries the minimum equations for $W$ can be
easily numerically integrated, yielding solutions qualitatively
independent from the physical dimension in the spherical case, and
strictly independent of the physical dimension in the planar case.
The main qualitative features can be obtained in the planar case
where the overlap profile can be obtained by quadrature. In fact
this case is formally identical to the problem of a 1D Newtonian
particle with coordinate $p$ evolving in time in a potential equal
to $-V(p)$. Unless otherwise stated the results we will discuss, and
in particular the behavior of the different lengths that will be
identified, are independent of the geometry and the dimension. We
will denote $p(x;\ell)$ the solution $p(x)$ of the variational
problem for the boundary conditions fixed at distance $\ell$.

\subsection{The physical picture}

The nature of the solutions of the stationary point of
(\ref{simple}) and the behavior of the PS correlations as a function
of the cavity size $\ell$ reflect the shape of the mean-field
potential $V(p)$ that we have discussed in the previous section.

In the ``very high'' temperature region $T>T^*$ for each value of
$\ell$, both in the spherical and planar geometries, there is a
unique solution of the field equations with the imposed boundary
conditions. As a function of $\ell$ the PS function $p(0;\ell)$
decreases continually from 1 to 0.

\begin{figure}[ht]
\begin{center}
\includegraphics[width= 0.7 \textwidth]{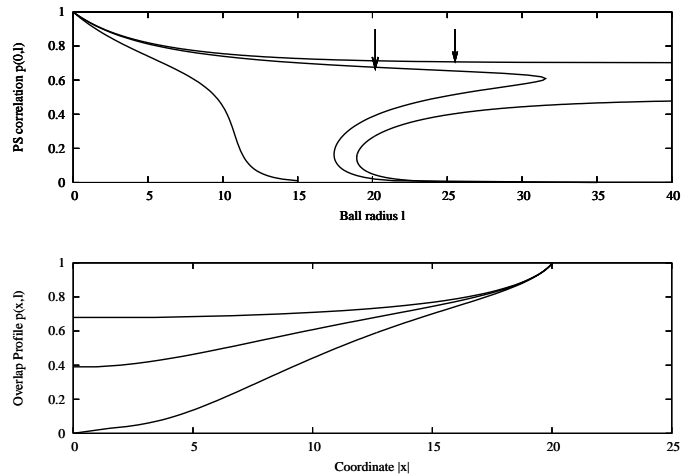}
\caption{Upper panel: values of the PS function $p(0,\ell)$ for
three values of the temperature in the different solutions:
$T=0.676> T^*=0.633$, $T=0.610$, in the interval $[T_{\rm
d},T^*]=[0.575,0.633]$ and $T=0.569$, in the interval
$[T_c,T_d]=[0.541,0.575]$. In the first case there is  a unique
solution to for all values of $\ell$. In the second case three
solutions are present for $\ell$ in the interval $[\ell_0,\ell_{\rm
d}]=[17.31,31.57]$. In the third case three solutions are present
for $\ell>\ell_0=19.47$. The arrow mark the points $\ell=\ell_{\rm
c}$. Below $\ell_{\rm c}$ the HOS is the thermodynamically favoured
state and the LOS is metastable, above that length the roles are
interchanged. For the $T=0.610$, $\ell_{\rm c}=20.4$, for $T=0.569$,
$\ell_{\rm c}=26.7$.
 Lower panel: The overlap profiles $p(x;\ell)$ corresponding to the
the three solutions for $\ell=20$ and $T=0.610$.  } \label{fig1}
\end{center}
\end{figure}

At $T=T^*$ an instability develops. Below that temperature one finds
a temperature dependent interval of lengths $I=[\ell_0,\ell_{\rm
d}]$ where the field equations admit three distinct solutions. For
temperatures in the interval $(T_{\rm d},T^*)$ both lengths are
finite, while $\ell_{\rm d}=\infty$ for $T<T_{\rm d}$. The behavior
of $p(0;\ell)$ for different temperatures is depicted in the figure
\ref{fig1}. The figure also shows the complete profile $p(x;\ell)$
as a function of $x$ for specific values of $\ell$ and the
temperature in the three solutions.
 According to the value of $p(0;\ell)$, we call the three
profiles Low Overlap Solution (LOS), Medium Overlap Solution (MOS)
and High Overlap Solution (HOS).  These have different
free-energies.  The HOS and the LOS one correspond to free-energy
absolute or relative minima, the MOS one is a maximum with respect
to $p(0;\ell)$.

The structure of the solutions allows to identify three relevant
lengths which grow for decreasing temperatures: two metastability
lengths $\ell_0$ and $\ell_{\rm d}$ defining the interval in which
three solutions exist, and a thermodynamic length $\ell_{\rm c}$,
with $\ell_0<\ell_{\rm c}<\ell_{\rm
  d}$, for which the HOS and the LOS have the same free-energy.  The
presence of two free-energy minima separated by a barrier naturally
interpreted a metastability phenomenon. The typical relaxation time
depends on size of the cavity $\ell$ and diverges exponentially for
$r_0\to\infty$ if $\ell_0< \ell< \ell_{\rm d}$.

We divide the remaining of this section in the discussion of the
equilibrium solutions, the metastable solutions and the unstable
solutions for $T<T^*$.

\subsubsection{Equilibrium and metastable solutions}
\label{herehere}

Let us consider the behavior of the system in spherical geometry as
a function of the ball radius $\ell$.  Below the length $\ell_0$ the
HOS is the only minimizer of the free-energy functional, there are
no metastable states poorly correlated with the boundary conditions.
This absence has consequences in dynamics: any configuration in $B$
chosen as initial condition for the dynamics, evolves in an $r_0$
independent time to the HOS.  Above this length, in the interval
$\ell\in(\ell_0,\ell_{\rm c})$ the LOS makes its appearance as a
free-energy local minimum. Though the HOS is still thermodynamically
dominant, metastable states exist such that if they are chosen as
initial conditions in relaxation dynamics, they will take a time
exponentially large in the interaction volume $r_0^d$ to relax to
equilibrium.

The difference in free-energy between the HOS and the LOS decreases
upon increasing $\ell$, until, for $\ell=\ell_{\rm c}$ the two
free-energy minima are degenerate. As we will discuss later, the
length $\ell_{\rm c}$ is growing for decreasing temperatures, and
diverges for $T\to T_{\rm c}$.  Above this length the LOS becomes
the dominating thermodynamic state.  There is a phase transition in
$\ell$ separating a phase strongly correlated with the boundary
below $\ell_{\rm c}$, from an essentially uncorrelated phase above.
The PS correlation function which should be identified with the
value of $p(0;\ell)$ in the thermodynamically dominant solution
exhibits a first order jump in $\ell=\ell_{\rm c}$. Below $\ell_{\rm
c}$ the system is thermodynamically in an ideal glassy amorphous
state where, despite no sign of intrinsic order, the boundary
conditions determine the bulk configuration.  Conversely, above
$\ell_{\rm c}$, the system is in a liquid situation where the
boundary has little influence on the bulk. The presence of the
transition is reminiscent of the analysis of PS function by Biroli
and Bouchaud~\shortcite{BiBo}, with $\ell_{\rm c}$ thus identified
with the mosaic length of phenomenological RFOT~\shortcite{KTW}.
This theory postulates a competition between interfacial energy and
configurational entropy, the contributions compensating exactly at
the transition.  However, as we have discussed for generic
observables in sec. \ref{sec_defpts_precise} (see formula
(\ref{obvious})), the average energy density for the constrained
system is equal in all points of space to its unconstrained
equilibrium value.  This implies in particular that at the
coexistence point both phases should have the same energy density,
and therefore entropy, in each point of space. The transition should
therefore depend on a different mechanism.
 Further analysis
of the HOS and the LOS presented below reveals that it has the
character of a configurational entropy crisis similar to the one
that happens at $T_c$ for the infinite system.

\begin{figure}[ht]
\begin{center}
\includegraphics[width= 0.7 \textwidth]{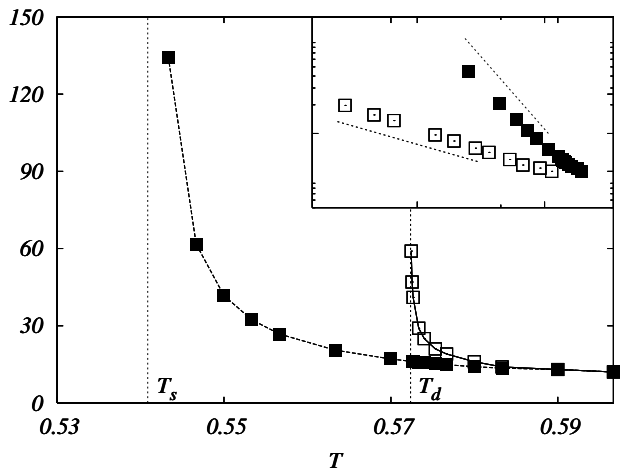}
\caption{Thermodynamic  (filled squares) and dynamic (empty squares)
lengths $\ell_{\rm c}$ and $\ell_{\rm d}$.  The vertical lines
correspond to the dynamic and thermodynamic glass transitions
$T_{\rm c}=0.541847$ and $T_{\rm d}=0.57525$.
 The lengths behave respectively as $\ell_{\rm c}\sim (T-T_{\rm c})^{-1}$ close to $T_{\rm c}$ and as $\ell_{\rm d}\sim (T-T_{\rm d})^{-1/4}$ close to $T_{\rm d}$.
 In the inset, the lengths as functions of  $(T-T_{\rm d})$  (for $\ell_{\rm d}$)
and $(T-T_{\rm c})$ (for $\ell_{\rm c}$). The dotted lines have
slope (respectively) −1/4 and −1.} \label{fig3}
\end{center}
\end{figure}

In the region $\ell\in(\ell_{\rm c},\ell_{\rm d})$, the HOS appears
to be a local minimum of the free-energy functional.  The length
$\ell_{\rm d}$ marks the limit of metastability of the HOS and was
called dynamical length in~\shortcite{FranzMontanari}.  For
$r_0\to\infty$ and $\ell\in (\ell_{\rm c},\ell_{\rm d})$, a system
prepared in the configuration $\us^{(0)}$ at an initial time $t=0$
would not be able to dynamically relax towards the thermodynamic
equilibrium profile.  For large but finite $r_0$ the relaxation time
(for the dynamics in a ball of radius $\ell$) to the LOS profile can
be expected to behave exponentially, $\tau(r_0,\ell,T)\sim
\exp(r_0^d {\mathcal B}(\ell, T))$. In other words, for
$r_0\to\infty$, while the relaxation time is divergent, the barrier
(counted in units of the interaction volume) $\lim_{r_0\to\infty}
\frac{1}{r_0^d} \log \tau(r_0,\ell,T)= {\mathcal B}(\ell, T)$
remains strictly positive. Conversely, for $\ell$ larger than
$\ell_{\rm d}$ the relaxation time would remain finite for $r_0\to
\infty$. $\ell_{\rm d}$ appears as the minimal length for which
relaxation is possible without barrier jumping. As Fig.~\ref{fig3}
shows, the dynamic length increases monotonically and diverges as
$T$ decreases from $T^*$ on approaching $T_{\rm d}$. Generically in
the class of models we are interested in, it behaves as
\begin{eqnarray}
\ell_{\rm d}\sim {(T-T_{\rm d})^{-\frac{ 1 }{4}}} \ ,
\end{eqnarray}
at least below dimension $6$. Interestingly,  this scaling, first
found in~\shortcite{KTW}, coincides with the one obtained for the
length $\ell_4$ quantifying the typical extension of collective
motion in MCT~\shortcite{FPlength,BBmct} relaxation processes. This
seems natural since, in the $r_0\to\infty$ limit a set of ideal
inhomogeneous MCT equations can be shown to describe the dynamics of
the Kac model~\shortcite{cortona}. A simple scaling argument
accounts for the above behavior in the present context. For
temperatures close to $T_{\rm d}$ and dimension smaller than
$d_c=6$, the properties of the solutions only depend on $V$ in the
vicinity of the inflection point~\shortcite{muratov}. The potential
can then be expanded to the cubic order around $p_{\rm d}$, its
inflection point at $T_{\rm d}$, as $V(p,T)=V(p_{\rm d},T)+ \epsilon
a (p-p_{\rm d})+b (p-p_{\rm d})^3$, where generically the
coefficient of the linear term is $\epsilon =T-T_{\rm d}$ while $a$
and $b$ are positive model-dependent parameters with a weaker
temperature dependence. Writing $p-p_{\rm d}=\varphi$ this gives
rise to a cubic field theory
\begin{eqnarray}
W[\varphi]=r_0^d \int d^d x \left(\frac c 2 (\nabla \varphi^2(x)) +a
\epsilon \varphi+b\varphi^3 \right) \ , \label{cubic}
\end{eqnarray}
whose critical properties dominate the behavior of observable
quantities for $\epsilon\to 0$.  In particular, equating the order
of magnitude of the different terms in the Lagrangian $\epsilon
\varphi \sim \varphi^3 \sim (\nabla \varphi)^2$ immediately leads to
a length $\ell_{\rm d}\sim \epsilon^{-1/4}$. It has been noticed in
the context of spinodal theories that the expansion (\ref{cubic})
cannot be used above dimension 6, and the critical point has a
non-universal character~\shortcite{muratov}.

\begin{figure}[ht]
\begin{center}
\includegraphics[width= 0.7 \textwidth]{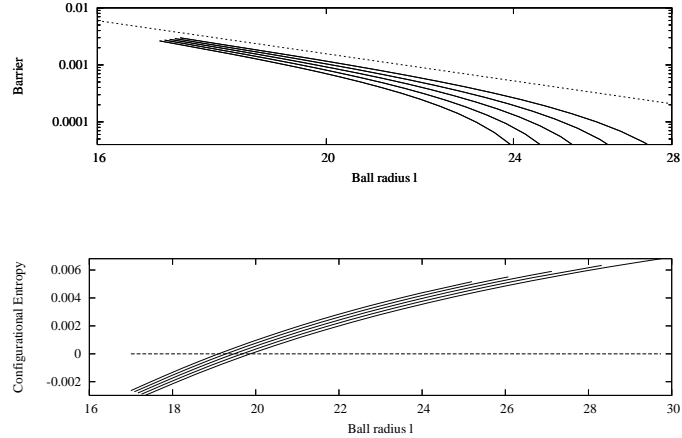}
\caption{Upper panel: The barrier density as a function of the ball
  radius $\ell$ for different temperatures. From bottom to top
  $T=0.5819, 0.5816, 0.5812, 0.5809, 0.5805$, all larger then
  $T_d=0.57525$. Lower temperatures correspond to higher barriers. The
  barrier goes to zero for $\ell\to\ell_{\rm d}$.  The dotted line a
  allows the comparison with the power law $\ell^{-6}$ expected from
  scaling close to $T_{\rm d}$. Lower panel: Configurational entropy
  density as a function of $\ell$ for the same temperatures as in the
  upper panel. The higher curves correspond to the higher
  temperatures. The configurational entropy curves touch zero at the
  mosaic length $\ell_{\rm c}$ and terminate at the dynamical length
  $\ell_{\rm d}$. Both lengths increase for decreasing
  temperatures. Below the length $\ell_{\rm d}$ the correct solution
  should include RSB.}
\label{fig:conf}
\end{center}
\end{figure}

The nature of the LOS is clarified considering the 1RSB solution for
$m=1$ (\ref{rsb1}) we mentioned in the previous section.  In the
region $\ell>\ell_{\rm d}$ the LOS is the unique free-energy
minimizer and it represents an ergodic state. Conversely, for
$\ell_{\rm c}<\ell<\ell_{\rm d}$ this solution describes a
non-ergodic state composed by a collection of ${\mathcal N}\sim
\exp(r_0^d \ell^d \Sigma(\ell,T))$ metastable states. The
free-energy difference between the HOS and the LOS, which, as
previously observed, should be purely entropic, in fact coincides
with ($T$ times) the configurational entropy in the constrained
system, denoted $\ell^d \Sigma(\ell,T)$. In other words there is an
exponentially large number $\mathcal N$ of metastable states
compatible with the boundary condition. The HOS just represents one
of them, on the same foot as all the others.  The behavior of
$\Sigma(\ell,T)$ is depicted in the lower panel of
Fig.~\ref{fig:conf}, which shows curves growing with $\ell$,
crossing the value 0 for $\ell=\ell_{\rm c}$. If $T>T_{\rm d}$ the
curves stop at the dynamical length $\ell_{\rm d}$ beyond which the
system recovers ergodicity.  For $T<T_{\rm d}$ the curves tend to
the bulk value of the mean-field configurational entropy
$\Sigma_\infty(T)$. The dependence of $\Sigma(\ell,T)$ upon $\ell$
can in both regions be described in a good approximation as
$\Sigma(T,\ell)=A(T)(1-\ell_{\rm c}/\ell)$, where for $T<T_{\rm d}$,
$A(T)=\Sigma_\infty(T)$. Upon decreasing the values of $\ell$
towards $\ell_{\rm c}$, the transition to the confined state is
driven by the reduction to zero of the configurational entropy. This
is decreased by a term that scales as $\ell^{d-1}$ relative to the
total configurational entropy $\ell^d \Sigma(\ell,T)$.

Let us mention without entering in a detailed discussion that in
addition to the solutions we discussed, low-overlap RSB solutions
with $m\ne 1$ (not respecting the conditional equilibrium
conditions) were also found in~\shortcite{FranzMontanari}. These
represent families of metastable states in the system with
free-energy higher then the equilibrium one.  For $\ell<\ell_{\rm
c}$ the solution with $m=1$ gives a negative configurational entropy
and is therefore inconsistent. The dominating metastable low overlap
state is one of the RSB $m\ne 1$ solutions just mentioned. It has
zero configurational entropy and positive energy difference with the
HOS.

The scenario for the various glass transitions as a function of
$\ell$ in the LOS resembles the one of the Mean-Field or diluted
$p$-spin models as a function of temperature.  We have an ergodic
phase for large $\ell>\ell_{\rm d}$, an intermediate phase with a
finite configurational entropy for $\ell_{\rm c}<\ell<\ell_{\rm d}$,
and a zero configurational entropy 1RSB phase at low $\ell<\ell_{\rm
c}$. We notice that the point at $\ell_d$ has the character of a
typical dynamical transition point of 1RSB systems. This has specifc
consequences on the dynamical properties (for the constrained
dynamics in a ball of radius $\ell$) when $\ell$ gets close to
$\ell_{\rm d}$, which should be similar to the one of the mean-field
model close to $T_{\rm d}$.  In particular, for $r_0\to\infty$, the
relaxation time grows and diverges if the system size goes to
$\ell_{\rm d}$ from above. On the same foot one can draw
consequences on the relaxation time within the metastable states
dominating below $\ell_{\rm d}$. One can argue that the stability of
these states decreases upon increasing $\ell$ until they become
marginal exactly at $\ell_d$~\shortcite{pspin-rev}. Accordingly, the
relaxation time within a metastable state diverges if $\ell$ tends
to $\ell_{\rm d}$ from below.

As already mentioned the dynamics of the model in the Kac limit is
exactly described by inhomogeneous MCT
equations~\shortcite{cortona}, so that the bulk relaxation time
behaves as $\tau(T)\sim \epsilon^{-\gamma}$ close to $T_{\rm d}$,
with a model dependent exponent $\gamma$~\shortcite{MCT}. For
$\ell>\ell_{\rm d}\sim \epsilon^{-1/4}$ one can expect the scaling
\begin{eqnarray}
\tau(\ell,T)=\frac{1}{\epsilon^\gamma} \hat{\tau}(\ell/\ell_{\rm d})
\ ,
\end{eqnarray}
where the scaling function $\hat{\tau}(u)$ tends to a constant at
large arguments, and has a power law behavior  $\hat{\tau}(u) \sim
(u-1)^{-\delta}$ for $u \gtrsim 1$. The exponent $\delta$ should be
in principle derived from the dynamical MCT-like
equations~\shortcite{cortona} and as $\gamma$ should depend on the
choice of the model. An analogous scaling should be expected below
$\ell_{\rm d}$ with $\tau(\ell,T)$ interpreted as the relaxation
time within a state, for the dynamics inside a ball of radius
$\ell$.

The thermodynamic mosaic length $\ell_{\rm c}$, whose temperature
behavior is shown in Fig.~\ref{fig3}, displays no singularity at
$T_{\rm d}$. This is coherent with the fact that for any finite
$r_0$ the dynamical transition is rounded-off and the relaxation
time remains finite, and is in agreement with the inequalities
(\ref{eq_bound}) of Sec.~\ref{relcorrtime}. Its behavior becomes
singular near the ideal glass transition point $T \gtrsim T_{\rm
  c}$. Here the two minima of $V$ are almost degenerate,  and $\ell_{\rm
  s}$ is large.  The solutions of the field equations in spherical
geometry for $d>1$ can be obtained through the ``thin wall
approximation'' of nucleation theory~\shortcite{langer}.  This
self-consistently neglects the thickness of the interface region,
where the solution passes form values of $p$ close to $q_{EA}$ to
values close to zero, this thickness being small with respect to
$\ell_{\rm c}$. Let us consider $\ell \gtrsim \ell_{\rm c}$. The
transition is driven by a bulk contribution to the configurational
entropy $\Sigma_\infty(T)\ell^d$ and an almost temperature
independent surface entropy reduction $-Y \ell^{d-1}$.  The value of
$Y$ can be computed in analogy with standard nucleation theory as
the action of the ``instantonic solution'' to the field equation
that connects the two degenerate minima of $V(p)$ at $T=T_{\rm c}$:
\begin{eqnarray}
Y=S_d \left. \int_0^{q_{EA}} dp \frac{2 V(p)} {\sqrt{
\frac{c}{2}\beta \phi''(p)V(p)} } \right|_{T=T_{\rm c}},
\end{eqnarray}
where $S_d$ is the surface of the unitary sphere in dimension $d$,
$S_d=2 \pi^{d/2}\Gamma(d/2)$. The length $\ell_{\rm c}$ is then
found from the cancellation of the total configuration entropy,
\begin{eqnarray}
\ell^d \Sigma(\ell,T)= \Sigma_\infty(T)\ell^d -Y  \ell^{d-1} \ ,
\label{totalentropy}
\end{eqnarray}
namely $\ell_{\rm c}=Y/\Sigma_\infty(T)$. The length $\ell_{\rm c}$
diverges at $T_{\rm c}$, with the scaling
\begin{eqnarray}
\ell_{\rm c}\sim \frac{Y}{\Sigma_\infty(T)}\sim \frac{Y}{T-T_{\rm
c}} \ . \label{kauz}
\end{eqnarray}
We will see in the next sub-section that this is also the typical
spatial extension of barrier states close to $T_{\rm c}$.  Notice
that the negative contribution to (\ref{totalentropy} ) is
proportional to the ball surface.  As we discuss in sec.
\ref{sec_fr0} phenomenological theories \shortcite{KTW,BiBo} suggest
that interface reduction terms could scale as $\ell^{\theta}$ with
$\theta<d-1$. As all qualitative results we present the value
$\theta=d-1$ can be checked to be independent on the choice of the
model in the 1RSB class or the various approximations like the thin
wall approximation or the gradient expansion of the interaction that
we have used to simplify the analytic treatment.  Our derivation,
however, depends crucially on the Kac limit, that in our analysis
preceeds the limit $T\to T_{\rm c}$. It is certainly possible that
for finite $r_0$ a non trivial cross-over could change this
exponent.

\subsubsection{The barrier state}

Let us complete our analysis with a discussion of the medium overlap
solution (MOS). In figures~\ref{fig1}  the value of $p(0;\ell)$
plotted in the MOS has been obtained supposing conditional
equilibrium as an approximation. Unfortunately, it is difficult to
go beyond this solution using numerical integration. Since by
definition the MOS is an unstable saddle of $W$, iterative methods
that work well for minima, simply do not find the solution.
Experience with the replica method shows however that such replica
symmetric solutions often give good approximations for the
free-energy and other quantities even in regions where the exact
description should include RSB.

We analyze below the regimes of temperatures close to $T_{\rm d}$
and $T_{\rm c}$ where scaling can be expected and analytic arguments
allow to go beyond numerical integration of the field equations. We
believe that our approximate solution while possibly failing in
numeric prefactors, gives back the correct scaling behavior
describing the dependence of the MOS on system size and temperature
for large $r_0$.

The MOS is interpreted physically as a dynamical barrier state to be
overcome during relaxation~\shortcite{barriere}.  The equilibrium
activation barrier ${\mathcal B}(\ell,T)$ can be estimated as the
free-energy difference between the  MOS and the HOS
\begin{eqnarray}
r_0^d {\mathcal B}(\ell,T)=W_{MOS}-W_{HOS} \ .
\end{eqnarray}
This is, for all temperatures, a decreasing function of the system
size $\ell$. For temperatures in the range $[T_{\rm d},T^*]$ the
barrier vanishes at $\ell_{\rm d}$, coherently activation is not
needed for relaxation in large systems. For temperatures below
$T_{\rm d}$ activation is required at all scales. The barrier for
bulk relaxation can be obtained as the large $\ell$ limit of
${\mathcal B}(\ell,T)$.

Close to $T_{\rm d}$ the behavior of the MOS and the HOS become
independent of the details of the model. Since we do not expect RSB
to affect the scaling, we can obtain the behavior of the various
quantities through dimensional analysis of the cubic field theory of
Eq.~(\ref{cubic}). The main results of this analysis are that
\begin{itemize}
\item For $T \gtrsim T_{\rm d}$ the barrier is non zero for $\ell<\ell_{\rm d}$ and
admits the scaling form
\begin{eqnarray}
{\mathcal B}(\ell,T)=\ell^{d-6}b_+(\ell /\ell_{\rm d}) \ ,
\end{eqnarray}
where the scaling function is such that the barrier vanishes
linearly as $\ell\to\ell_{\rm d}$: $b_+(x)\approx Const. \times
(1-x)$ for $x\to 1$ and $b_+(x)=0$ for $x>1$. The behavior of the
barrier in this region is shown in the upper panel of
fig.~\ref{fig:conf}.
\item For $T \lesssim T_{\rm d}$ the barrier admits the scaling form
\begin{eqnarray}
{\mathcal B}(\ell,T)=|\epsilon|^{\frac{6-d}{4} }b_-(\ell
|\epsilon|^{1/4}) \ , \label{scaba}
\end{eqnarray}
where the function $b_-(x)$ is a decreasing function of $x$ that
admit a positive limit for $x\to\infty$.  As already noticed, from
(\ref{scaba}) we can obtain the barrier for relaxation in the bulk
as the limit of large $\ell$ of ${\mathcal B}(\ell,T)$. This
coincides with a direct estimate of the barrier
in~\shortcite{SilvioJSTAT,D0} where boundary conditions are imposed
at infinity and scales as ${\mathcal B}_\infty(T)\sim
|\epsilon|^{\frac{6-d}{4} }$.  The scaling (\ref{scaba}) shows that
the asymptotic value is reached on lengths of order $\ell_{\rm
b}\sim|\epsilon|^{-1/4}$.  This is then the typical spacial
extension of dominating barrier modes governing the relaxation in
the bulk, which becomes large on approaching $T_{\rm d}$ from below.
\end{itemize}

Let us now turn to temperatures close to the static transition
temperature $T_c$. Analogously to the case of the equilibrium
solutions of the previous subsection, for small $T-T_{\rm c}$
nucleation arguments can be used to study the barrier. For large
$\ell$, the barrier solution $p(x;\ell)$ has the shape of a bubble
of small overlap values in a sea where $q(x;\ell)\approx q_{EA}$.
This is surrounded by a crown around the boundary of $\Bbar$ where
$p$ decays from 1 to $\approx q_{EA}$. For $d>1$ the  typical bubble
radius  $\ell_{\rm b}$, and the barrier value can be just estimated
from the point of maximum of expression
(\ref{totalentropy})\footnote{The 1d case requires a special
analysis and one finds that $\ell_{\rm b}\sim -\log(T-T_{\rm c})$.}.
This gives, as announced, a common scaling for  $\ell_{\rm b}$ and
$\ell_{\rm c}$,
\begin{eqnarray}
\ell_{\rm b}\sim \frac{Y}{\Sigma_\infty(T)} \sim  (T-T_c)^{-1} \ ,
\end{eqnarray}
and for the barrier the behavior
\begin{eqnarray}
{\mathcal B}(\ell,T)\sim Y^d (\Sigma_\infty(T))^{-(d-1)}\sim
(T-T_{\rm c})^{-(d-1)} \ , \label{vogelfulcher}
\end{eqnarray}
which is divergent for $d>1$ when $T \to T_{\rm c}$. Eq.
(\ref{vogelfulcher})  has the form of a modified Adams-Gibbs inverse
proportionality relation between the relaxation barrier and the
configurational entropy\shortcite{Adam}. In turn, using
$\Sigma_\infty(T)\sim T-T_{\rm c}$, one finds a modified
Vogel-Fulcher law\footnote{The usual Vogel-Fulcher form posits a
barrier
  proportional to $(T-T_{\rm c})^{-1}$.}  with an exponent that
depends on dimensionality, with associated lower critical dimension
$d_{LCD}=1$.  We remark the coincidence of this result in the Kac
limit with the naive analysis of~\shortcite{KTW}. The comment we
made at the end of section (\ref{herehere}) on  the exponent can be
verbatim applied here to the modified Vogel-Fulcher exponent.

\subsubsection{Correlations in wall geometry}

Before concluding our discussion on the Kac limit, let's have a
closer look to the  three solutions for large $\ell\gg \ell_{\rm c}$
and $T \gtrsim T_{\rm c}$. The large $\ell$ limit is equivalent to
imposing the boundary condition on a single planar wall. This
condition has been object of recent numerical
experiments~\shortcite{KobPS} and deserves a short comment.

The question here is up to which distance the boundary has an effect
on the configuration of the system, i.e. what is the typical
distance from the boundary over which $p(x;\ell)$ decays to zero and
if this distance coincides with $\ell_{\rm c}$. The behavior of the
three solutions in this regime can be easily worked out from the
field equation. The metastable HOS never looses correlations: after
a rapid decay from 1 on the wall, which is common with the MOS and
the LOS, stabilizes to $p(x;\ell)\simeq q_{EA}$. The unstable MOS
persists to values $p(x;\ell)\approx q_{EA}$ up to distances of
order $\ell$ from the wall. It decays to zero at a distance
$\ell_{\rm c}$ from the center and is similar to a ball of radius
$\ell_{\rm c}$ of small overlap in a sea of large overlap values.
More interesting is the behavior of the LOS which gives the
equilibrium profile.  In this case the overlap decays to low values
over distances of order of a ``wetting'' length $\ell_{\rm
w}\sim-\log(T-T_{\rm c})$ from the boundary before decaying to
values close to zero. We conclude that the semi-infinite wall
geometry fails to identify the correlations $\ell_{\rm c}$ while it
identifies a much smaller length $\ell_{ \rm w}$.

% \begin{figure}[ht]
% \begin{center}
% \includegraphics[width= 0.7 \textwidth]{wall.eps}
% \caption{Upper panel: the three solutions for $T= 0.541859$
% ($T_{\rm c}=0.541847$).
% The decay to $q_{EA}$ is common to the three solutions. The functional form
% of the decay from  $q_{EA}$ is the same in the LOS and in the MOS. The
% LOS decays at distances of order $-\log(T-T_{\rm c})$ from the boundary. The
% MOS decays at distances of order $(T-T_{\rm c})^{-1}$ form the center.
% Lower panel: the LOS for $T= 0.541859, 0.541856, 0.541853, 0.54185$ (Lower temperatures to the right). }
% \label{fig:wall}
% \end{center}
% \end{figure}

\subsubsection{Beyond the Kac limit}
\label{sec_fr0}

The results presented so far are strictly valid in the Kac limit.
We briefly discuss here some possible scenarios for $r_0$ large but
finite.

We would like first to discuss the rounding off of the dynamical
(MCT) transition. The analysis can be performed in analogy with
ferromagnetic models with Kac interactions below the lower critical
dimension where ``finite range scaling'' (FRS)
holds~\shortcite{binder}. The basic hypothesis of FRS is that the
interaction range acts as a cut-off to critical behavior in a
similar way as finite size does for usual critical phenomena. This
idea has been employed in a 1d Kac spin glass with a continuous
transition in the Kac limit, to predict the rate of growth of the
correlation length with $r_0$ at the mean-field critical temperature
of the model~\shortcite{FPSG1D}. The $r_0\to\infty$ dynamical
critical point has a mean-field character, so that one can expect
mean-field scaling. An observable $O$ which depends on $T$, $r_0$
and $\ell$, should be described by scaling as:
\begin{eqnarray}
O(T,r_0,\ell) = \epsilon^{-y_O}\hat{O}(r_0^{a}\epsilon^{1/4}, \ell
\epsilon^{1/4}) \ .
\end{eqnarray}
The scaling function $\hat{O}$ is such to cut-off both the
singularity that appear for $r_o\to\infty$ at $\ell=\ell_{\rm d}$
for $\epsilon>0$ and the one at $\ell\to\infty$ and $\epsilon=0$.
The value of the exponent $a$ could be guessed if one used the
potential functional as an ordinary field theory beyond the Kac
limit for which it has been derived. Simple rescaling of the cubic
action (\ref{cubic}) leads then to $a=\frac{d}{6-d}$, implying that
the cross-over from Mean-Field MCT behavior to non-critical behavior
happens on length scales of the order of
$\ell_1=r_0^{\frac{d}{6-d}}$. The same conclusion could be reached
analysing the behavior of the barrier for $T \lesssim T_{\rm
  d}$ and $\ell \lesssim \ell_{\rm d}$, which scales as $r_0^d
{\mathcal B}(\ell,T)\sim r_0^d \ell^{d-6}(1-\ell/\ell_{\rm d})$.
This analysis, however, relies on a continuum approximation of the
potential $W[p(x)]$, which for finite $r_0$ neglects spatial
fluctuations of the quenched disorder and heterogeneities in the
reference configuration $\us^{(0)}$.  It has been recently observed
that finite size scaling around mean-field critical points in
disordered systems can be non-trivial due to effective fluctuations
of the critical temperature~\shortcite{BiroliFSS,FPRR}.
 Analogously here we can expect that effective
local fluctuations of the critical temperatures can affect FRS. A
criterion for the validity of naive FRS can be obtained in analogy
with the analysis of FSS in disordered systems which in turn is an
application of Harris criterion \shortcite{Harris} for relevance of
disorder in phase transitions.

Let us rewrite the naive FRS of the barrier for temperatures
%free-energy scaling
slightly smaller than $T_{\rm d}$, which for $\ell\to\infty$ we
write as
\begin{eqnarray}
r_0^d {\mathcal B}_\infty(T) \sim r_0^d |\epsilon|^{\nu_{MF}
(d_c-d)}
%\hat{F}(\ell \epsilon^{1/4}).
\label{frisca}
\end{eqnarray}
%ATTENZIONE a $L$.
In our case, $\nu_{MF}=1/4$ and $d_c=6$; the form (\ref{frisca})
offers the possibility of studying more general cases. The scaling
(\ref{frisca}) gives a characteristic cross-over length $\ell_1 \sim
r_0^{\frac{d}{d_c-d}}$ beyond which deviations from mean-field
theory are to be expected (remember that lengths are measured in
units of $r_0$). One can expect violations of this naive FRS when,
on scale $\ell_1$ the typical fluctuations of the critical
temperature are larger then the average deviation $|\epsilon|$. The
critical temperature fluctuations, related to local disorder
fluctuations, can be expected to be of the order $(r_0
\ell_1)^{-d/2}$. Applying the above criterion, we find that naive
FRS is violated if $r_0^{-(\frac{d}{(d_c-d)\nu_{MF}})} \ll r_0
^{-\frac d 2 \frac{d_c}{(d_c-d)}}$ i.e. if $d_c \nu_{MF}<2$. It is
notable that this relation coincides with the criterion for the
validity of naive FSS for disordered systems above the upper
critical dimension.  In our case $d_c \nu_{MF}=3/2$ and so that
naive scaling cannot not be expected to hold. In order to understand
the nature of the cross-over for finite $r_0$ a deep analysis of the
effect of critical temperature fluctuations is needed
\shortcite{FPRR}. This goes beyond the scope of the present review
and won't be attempted here.

The second point we would like to discuss is the behavior of the
barrier close to $T_{\rm c}$.  The Kac limit prediction is that the
configurational entropy reduction in a finite volume of linear size
$\ell>\ell_{\rm c}$ is proportional to the surface, $\delta \Sigma
\ell^d=-Y\ell^{d-1}$, which, as observed, leads to a modified
Vogel-Fulcher law for the relaxation time $\tau\sim \exp(r_0^d
(T-T_{\rm c})^{-(d-1)})$.  Of course this does not necessarily imply
an ideal glass transition if $r_0$ is finite. It just tells that for
growing $r_0$ and fixed temperature $T\leq T_{\rm c}$ the relaxation
time grows faster then $\exp(r_0^d {\mathcal B})$ for any ${\mathcal
B}>0$. The result is compatible both with the absence of a
singularity at finite temperature or with a singularity that could
be weaker than the one found in the Kac limit. In phenomenological
RFOT it is supposed that the interface exponent is renormalized by
the effect of fluctuations, yielding $\delta \Sigma
\ell^d=-Y\ell^{\theta}$ with $\theta\leq d-1$. In
ref.~\shortcite{KTW}, scaling arguments in favor of the value
$\theta=d/2$ have been put forward, which leads to the canonical
Vogel-Fulcher form in all dimensions $d\geq 2$. Correspondingly, the
static length $\ell_{\rm c}$ would behave as $(T-T_{\rm c})^{-d/2}$
rather than $(T-T_{\rm c})^{-1}$ as found for $r_0\to\infty$.  If
these suggestions are correct, for large but finite $r_0$ there
should be a cross-over between the two regimes. A Ginzburg criterion
involving $r_0$ as well as $T-T_{\rm c}$ should govern this
cross-over\footnote{Note that if we ask here that the fluctuations
of the critical temperature on a correlated volume are smaller than
$T-T_{\rm c}$ we get the condition $(T-T_{\rm c}) \gg \frac{1}{(r_0
\ell_{\rm c})^{d/2}}$ which is always verified for $\ell_{\rm c}
\sim (T-T_{\rm c})^{-1}$.}. Differently from the cross-over close to
$T_{\rm d}$ the formalism we have used does not seem to suggest a
mechanism for this cut-off.

\section{Conclusions}

This chapter concerns the inclusion of spatial aspects in glassy
theory. In particular the emphasis has been put on recently proposed
measures of correlations -the point-to-set correlations- that
involve an infinite number of variables.

In the first part of the chapter we reviewed exact general bounds
that relate the growth of a relaxation time to the growth of a
correlation length.  We first introduced with a progressive
mathematical rigor the idea of point-to-set correlation functions
and the associated correlation lengths, and explained to which
extent this definition allows to reconcile the phenomenology of
glasses with the intuitive association between growing time and
length scales. These bounds do not spoil the interest of studying PS
correlations in glassy systems. The lower and upper bounds have
different forms, which correspond to different physical mechanisms
of relaxation. The lower-bound is, up to a non-trivial exponent, of
the critical dynamics type, while the upper-bound has the form of
activated dynamics.

In the second part of this chapter we discussed this correlation
length in the context of the mean-field picture of glassy phenomena
provided by 1RSB disordered models. We have thus reviewed some
analytical works dedicated to growing length scales in simplified
models of glasses and their puzzling relationship with growing time
scales. We discussed diluted random graph models and Kac like models
which add new elements to the already rich behaviour of
fully-connected models of the $p$-spin family.  For random graph
models we found that the divergence of  the point-to-set length
accompanies the one of the relaxation time at the dynamical
transition.

The study of the Kac model reveals a rich scenario with two relevant
correlation lengths, the mosaic length below which a system behaves
thermodynamically as an ideal glass, and a dynamical length
quantifying the typical extension of relaxation modes which do not
require activation. At the dynamical transition the former remain
finite, coherently with the fact that for any finite interaction
range the transition is rounded-off, while the latter diverges.

Let us close this discussion with a few open questions.

The upper bound in Eq.~(\ref{eq_bound}) implies that the divergence
of the equilibrium correlation time is necessarily accompanied by a
divergence of the point-to-set correlation length, even though with
a possibly much slower form of the divergence. Both in diluted
random graph models, and in spin models in the Kac limit, the
divergence of the point-to-set correlation length at $T_{\rm d}$
cannot be detected by the study of static $n$-point correlation
function, for any finite $n$. Hence in these models there is a true
separation between the dynamic transition and the thermodynamic one,
the free-energy having a singularity only at a lower temperature
$T_{\rm c}$. An important open question is to determine whether this
phenomenon of growth of the point-to-set correlation function
without a trace in the two-point function can persist in
finite-dimensional models or if this is an artifact of mean-field
models. In the latter case it would mean that bounds of the form
(\ref{eq_bound}) holds with $\ell$ being replaced by a standard
two-point correlation length (and different constants $C_{1,2}$).
This should be important from a theoretical point of view, yet would
not contradict the experimental and numerical situation. On the
range of correlation time where measurements are possible the growth
of the static correlations implied by this hypothetical extension of
(\ref{eq_bound}) could still be very weak.

Other questions relate to the study of finite range disordered
models beyond the Kac limit. An accomplished theory of finite range
scaling close to the dynamical temperature $T_d$ should enable to
theoretically describe the crossover between mode coupling regime to
activated dynamics. A related question is the connection between the
purely static correlation length discussed here and the dynamical
one extracted from four-points (two-time, two-location) dynamical
susceptibilities. As we have discussed, the dynamical length found
in MCT above $T_d$ can be detected in the Kac limit analyzing
metastable solutions of the field equations in finite geometry. In
this case, dynamic and static correlation length do not coincide.
It would be important to better understand the quantitative
relationship between these two characterizations of the correlations
in a system.

\bigskip

\centerline{\bf Acknowledgments}

\medskip

We warmly thank Andrea Montanari, Giorgio Parisi and Fabio-Lucio
Toninelli for fruitful collaborations on several works presented
here.

\end{document}